\documentclass[aps,prev,onecolumn,preprintnumbers,floatfix,nofootinbib]{revtex4-1}
\pdfoutput=1
\usepackage{geometry,amsmath,amsfonts}
\usepackage{slashed}
\usepackage{epsfig}
\usepackage{latexsym}
\usepackage{graphicx}
\usepackage[justification=RaggedRight]{caption}
\usepackage{amssymb}
\usepackage{subfig}
\usepackage{color}
\usepackage{multirow}
\usepackage{color}
\usepackage{rotating}
\usepackage{ifthen}
\usepackage{epsfig}

\usepackage[utf8]{inputenc} 
\usepackage[T1]{fontenc}

\begin{document}

\title{Beyond the Veil: Charting WIMP Territories at the Neutrino Floor}
\author{Giorgio Arcadi$^{a,b}$}
\email{giorgio.arcadi@unime.it}
\author{Manfred Lindner$^{c}$}
\email{lindner@mpi-hd.mpg.de}
\author{Stefano Profumo$^{d,e}$}
\email{profumo@ucsc.edu}

\vspace{0.1cm}
 \affiliation{
${}^a$ 
 Dipartimento di Scienze Matematiche e Informatiche, Scienze Fisiche e Scienze della Terra, Universita degli Studi di Messina, Viale Ferdinando Stagno d'Alcontres 31, I-98166 Messina, Italy
}

\vspace{0.1cm}
 \affiliation{
${}^b$ 
INFN Sezione di Catania, Via Santa Sofia 64, I-95123 Catania, Italy
}

\vspace{0.1cm}
\affiliation{${}^c$
Max Planck Institut f\"ur Kernphysik, Saupfercheckweg 1, D-69117 Heidelberg, Germany}

\vspace{0.1cm}
\affiliation{${}^d$ Department of Physics, University of California, Santa Cruz, 
1156 High St, Santa Cruz, CA 95060, United States of America}
\affiliation{${}^e$ Santa Cruz Institute for Particle Physics, University of California, Santa Cruz, 
1156 High St, Santa Cruz, CA 95060, United States of America}

\begin{abstract} 
We establish comprehensive theoretical benchmarks for Weakly Interacting Massive Particles (WIMPs) accessible to ultimate direct detection experiments, focusing on the challenging parameter space between current experimental limits and the irreducible neutrino background. We systematically examine both thermal freeze-out and freeze-in production mechanisms across a range of simplified dark matter models, including $s$-channel scalar and vector portals, $t$-channel mediator scenarios, and electroweakly interacting multiplets. For thermal relics, we identify parameter regions where suppressed direct detection cross-sections naturally arise through momentum-dependent interactions and blind-spot configurations, while maintaining the correct relic abundance. We extensively investigate freeze-in scenarios, demonstrating how feebly interacting massive particles (FIMPs) in portal models can populate experimentally accessible parameter space despite their ultra-weak couplings. Additionally, we explore how non-standard cosmological histories—including early matter domination and fast-expanding Universe scenarios—can dramatically alter the relationship between relic density and detection prospects, opening new avenues for discovery. Our analysis provides a roadmap for next-generation experiments approaching the neutrino floor, highlighting complementary detection strategies and identifying the most promising theoretical targets for ultimate sensitivity dark matter searches. These benchmarks establish the theoretical foundation for the final push toward comprehensive coverage of well-motivated WIMP parameter space.
\end{abstract}

\maketitle

\section{Introduction}
The nature of dark matter remains one of the most profound puzzles in modern physics, with compelling evidence from cosmological observations indicating that approximately 85\% of the matter content of the Universe consists of non-luminous, non-baryonic particles. Among the leading theoretical candidates, Weakly Interacting Massive Particles (WIMPs) have emerged as particularly well-motivated due to the remarkable coincidence that particles with weak-scale masses and interactions naturally produce the observed dark matter relic density through thermal freeze-out—a phenomenon often referred to as the ``WIMP miracle''. This mechanism predicts that dark matter particles were initially in thermal equilibrium with Standard Model particles in the early Universe, subsequently decoupling when their annihilation rate fell below the Hubble expansion rate, leaving behind a relic abundance that matches observations for cross-sections of order \(\langle\sigma v\rangle \sim 3 \times 10^{-26}\) cm\(^3\)/s.

Direct detection experiments have made extraordinary progress in probing WIMP-nucleon interactions, with successive generations of increasingly sensitive detectors pushing the experimental frontier toward ever-smaller cross-sections~\cite{Billard:2013cfa,Akerib:2022ort}. Current world-leading experiments such as LUX-ZEPLIN (LZ) and XENONnT have achieved sensitivities approaching \(10^{-47}\) cm\(^2\) for spin-independent WIMP-nucleon scattering, while next-generation multi-ton scale detectors like DARWIN/XLZD promise to extend this reach by another order of magnitude. However, these experiments are rapidly approaching a fundamental limit known as the neutrino floor or neutrino fog, where coherent elastic neutrino-nucleus scattering (CEvNS) from astrophysical sources creates an irreducible background that mimics the expected WIMP signal~\cite{Dent:2016wcr,OHare:2020wah,Sierra:2017pvt}.

The impending approach to the neutrino floor necessitates a comprehensive theoretical framework to identify the most promising WIMP scenarios that remain accessible to ultimate direct detection experiments. This challenge is compounded by the recognition that traditional thermal freeze-out models, while elegantly simple, represent only one possible mechanism for dark matter production. Alternative scenarios, including freeze-in production of feebly interacting massive particles (FIMPs)~\cite{Arcadi:2024def,Hambye:2018dpi,Cosme:2023bkx} and dark matter generation in non-standard cosmological histories~\cite{Arias:2019abc,SilvaMalpartida:2024abc}, can dramatically alter the relationship between relic abundance and direct detection cross-sections, potentially opening new windows for discovery.

Simplified dark matter models have emerged as powerful theoretical tools for systematic exploration of the WIMP parameter space, providing model-independent frameworks that capture the essential physics while avoiding the complexities of full ultraviolet completions~\cite{Kahlhoefer:2015vua}. These models typically feature a dark matter candidate coupled to Standard Model particles through various portal interactions, including scalar (Higgs portal), vector (dark photon), and fermion (sterile neutrino) mediators. Such frameworks enable systematic mapping of the complementarity between different experimental approaches—direct detection, indirect detection, and collider searches—while maintaining theoretical rigor and predictive power.

The theoretical landscape becomes considerably richer when considering departures from the standard thermal freeze-out paradigm. Freeze-in scenarios, where dark matter is produced through the extremely feeble interactions of particles that never achieve thermal equilibrium, naturally accommodate ultra-small couplings that would be challenging to probe experimentally~\cite{Belanger:2023whi,Heeba:2023rjq}. These models predict distinctive relationships between production mechanisms and detection prospects, often requiring dedicated analysis techniques to properly account for the non-thermal nature of the dark matter distribution. Similarly, non-standard cosmological histories—such as early matter domination phases or periods of accelerated expansion—can significantly modify the freeze-out dynamics, leading to viable dark matter candidates with properties that differ substantially from standard expectations~\cite{DEramoFernandez:2017abc,DEramoFernandez:2018def}.

Electroweak multiplet dark matter models represent a particularly compelling class of WIMP candidates, as they emerge naturally from extensions of the Standard Model gauge structure and provide concrete predictions with minimal free parameters~\cite{Cirelli:2005uq,Cirelli:2007xd}. The so-called ``minimal dark matter'' scenarios, featuring pure SU(2) multiplets with specific hypercharge assignments, offer fully predictive frameworks where the dark matter mass and interaction strength are determined by the electroweak quantum numbers alone~\cite{Hisano:2011cs,Cirelli:2014nga}. These models exemplify the type of well-motivated theoretical targets that next-generation experiments should prioritize.

Recent developments in both experimental techniques and theoretical understanding have highlighted the critical importance of loop-induced effects and radiative corrections in determining direct detection prospects. Many theoretically attractive scenarios feature suppressed tree-level interactions with nucleons, either through spin/momentum dependencies or accidental cancellations, making loop contributions the dominant source of detectable signals~\cite{Arcadi:2017jse}. Understanding these quantum corrections is essential for making reliable predictions about the ultimate reach of direct detection experiments.

The approach to the neutrino floor also motivates exploration of complementary experimental strategies, including directional detection capabilities that can potentially distinguish between neutrino-induced and WIMP-induced nuclear recoils based on the expected anisotropy of the WIMP signal~\cite{Gelmini:2018gkh,Boehm:2018luh}. Multi-target approaches, employing different detector materials with varying response functions, offer another promising avenue for breaking the degeneracy between neutrino backgrounds and potential WIMP signals~\cite{Chao:2019bqg,Davis:2014cla}.

In this work, we establish comprehensive theoretical benchmarks for WIMP dark matter that remain accessible to ultimate direct detection experiments operating at or near the neutrino floor. We systematically examine representative simplified models spanning the major portal scenarios, carefully analyzing both thermal and non-thermal production mechanisms while accounting for the full range of theoretical uncertainties. Our analysis incorporates the latest constraints from current experiments while projecting sensitivity reaches for next-generation detectors, providing a roadmap for the final phase of the direct detection experimental program. Through detailed investigation of non-standard cosmological scenarios and careful treatment of radiative corrections, we identify the most promising theoretical targets for ultimate sensitivity searches and establish the foundations for interpreting results from experiments approaching the fundamental limits of current detection technology.


\section{Benchmark Models}

The theoretical landscape of dark matter (DM) phenomenology has evolved to embrace a diverse array of production mechanisms beyond the traditional thermal freeze-out paradigm. While freeze-out remains a compelling framework due to its natural connection between annihilation and relic abundance, the increasingly stringent constraints from direct detection experiments have motivated the exploration of alternative scenarios. Among these, freeze-in production mechanisms have gained particular attention, offering potentially viable pathways to the observed DM relic density while maintaining compatibility with experimental bounds.

Portal models represent a particularly elegant framework for investigating these diverse production scenarios. By connecting the dark sector to the Standard Model through well-motivated mediators, these models provide a systematic approach to understanding the interplay between DM production, relic density constraints, and detection prospects. The simplicity of portal interactions allows for analytical insights while maintaining sufficient complexity to capture the essential physics of DM phenomenology.

In this section, we examine several representative portal models, each illustrating distinct aspects of the relationship between DM production mechanisms and experimental signatures. Our analysis encompasses both scalar and fermionic DM candidates, coupled through various mediator types, providing a comprehensive survey of the theoretical possibilities. We pay particular attention to the tension between achieving the correct relic abundance through thermal freeze-out and avoiding exclusion by direct detection experiments, demonstrating how this tension motivates alternative production mechanisms.

As already pointed out, portal models allow us to exploit different DM production scenarios, ranging from the conventional freeze-out paradigm to freeze-in mechanisms. We consider several relevant examples of these scenarios, each characterized by a distinctive interplay between DM relic density requirements and detection strategies:

\begin{itemize}
 \item {\it $s$-channel simplified model scalar DM}: This model considers a real scalar SU(2) singlet DM candidate\footnote{We focus here on a real scalar, motivated by simplicity. The case of a complex DM candidate would present only negligible differences.} interacting with a spin-0 BSM state $S$ according to the following lagrangian:
\begin{equation}
\mathcal{L}=\mu_\chi^S \lambda_\chi^S \chi \chi S 
+\frac{1}{2} (\lambda_\chi^S)^2 \chi^2 S^2+\frac{c_S}{\sqrt{2}}\frac{m_f}{v_h}\overline{f} f S
\end{equation}
The most minimal realization of this scenario is the so-called Higgs portal model, where $S \equiv h$ and $c_S=1,\mu_\chi^S=v_h$ with $v_h=246\,\mbox{GeV}$ being the Higgs vacuum expectation value. The designation ``$s$-channel portal'' stems from the fact that DM particles can annihilate into SM fermion pairs via $s$-channel exchange of the mediator field $S$.

\item {\it $s$-channel simplified model fermion DM}: This model is analogous to the previous case but with fermionic DM\footnote{There are no substantial differences between Majorana and Dirac fermions. We focus on the latter case for simplicity and definiteness.}:
\begin{equation}
\mathcal{L}=g_\psi  \psi \psi S+\frac{c_S}{\sqrt{2}}\frac{m_f}{v_h}\overline{f} f S,\nonumber\\
\end{equation}

\item {\it Simplified vector portal}: We consider a Dirac fermion coupled vectorially with a spin-1 $s$-channel mediator, denoted $Z^{'}$, which is in turn coupled vectorially to SM fermion pairs:
\begin{equation}
\mathcal{L}=g_\psi^{V} \overline{\chi} \gamma^\mu \chi Z_\mu^{'}+g_f^V \overline{f} \gamma^\mu  f Z_\mu^{'}.
\end{equation}
This model has been selected as it represents the most constrained scenario in conventional thermal freeze-out; consequently, it is particularly interesting to consider alternative DM production mechanisms in this context. Furthermore, the choice of purely vectorial couplings is motivated by the fact that they are not affected by issues related to perturbative unitarity \cite{Kahlhoefer:2015bea}. Finally, this model serves as a popular benchmark for light DM searches.

\item {\it Simplified $t$-channel model}: We consider a Majorana fermion with a Yukawa-type interaction with a BSM scalar $\Phi$ and a SM fermion:
\begin{equation}
    \mathcal{L}=
\lambda_\chi \overline{f}{P_R}{\Phi}\chi+{\rm{h.c.}}
\end{equation}
The ``$t$-channel portal'' designation comes from the fact that the interaction above generates DM annihilations into SM fermions via $t$-channel exchange of the mediator. $t$-channel portals can be realized for different combinations of DM/mediator spins (see e.g. \cite{Arina:2020tuw,Arcadi:2021glq,Arcadi:2021cwg,Arina:2023msd,Arcadi:2023imv,Arina:2025zpi} for more comprehensive studies). The choice of a Majorana fermion DM and a $t$-channel scalar mediator is motivated by the fact that this model is characterized by $p$-wave dominated DM annihilation cross-sections and, most importantly, loop-induced spin-independent (SI) interactions between DM and nucleons.
\end{itemize}

In addition to the aforementioned portals, we consider one of the most popular archetypes for WIMP DM, namely electroweakly interacting DM. Focusing on the case of a Majorana fermion, we consider the following lagrangian \cite{Hisano:2011cs}:
\begin{equation}
    \mathcal{L}=\frac{g_2}{4}\sqrt{n^2-(2Y+1)^2}\overline{\chi}\gamma^\mu \psi^{-}W_\mu^{+}+\frac{g_2}{4}\sqrt{n^2-(2Y-1)^2}\overline{\chi}\gamma^\mu \psi^{+}W_\mu^{-}-\frac{ig_2 Y}{\cos\theta_W}\overline{\chi}\gamma^\mu \eta Z_\mu
\end{equation}
Here, $n$ represents the type of SU(2) multiplet the DM belongs to and $Y$ the corresponding hypercharge. The so-called minimal DM models \cite{Cirelli:2005uq,Cirelli:2007xd} fall into this category, as for suitable assignments of $(Y,n)$ the DM can be made cosmologically stable without the introduction of ad hoc symmetries. For our study we will consider the case $n=2,Y=1/2$ (analogous to the supersymmetric higgsino). Other interesting cases are $n=3,Y=0$, corresponding to pure Wino DM in supersymmetric models \cite{Cirelli:2014dsa}, and $n=5,Y=0$, corresponding to one of the original minimal DM proposals.

\begin{figure}
    \centering
    \includegraphics[width=0.5\linewidth]{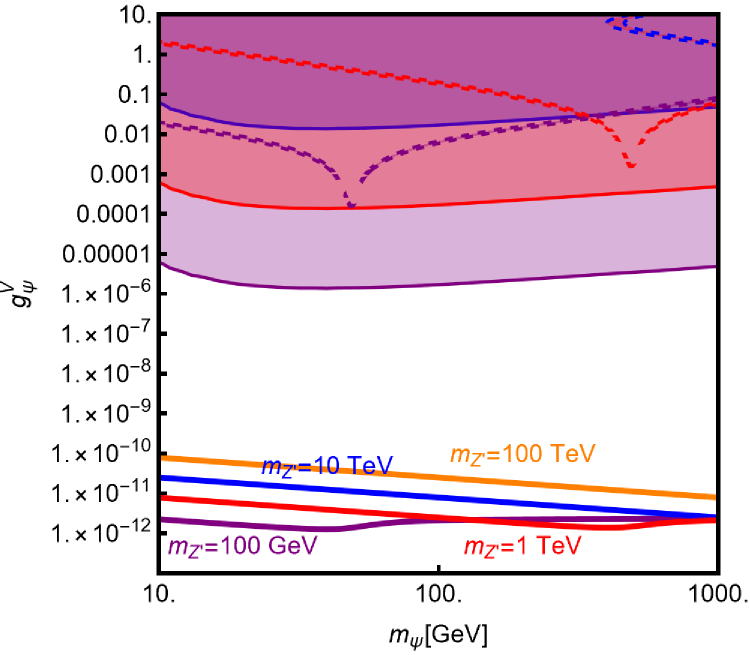}
    \caption{\footnotesize{{\it Dashed lines} Isocontours of the correct relic density, for the spin-1 mediator simplified model, in the $(m_\psi,g_\psi^V)$ bidimensional plane, assuming the thermal freeze-out paradigm. The different colors correspond to different values, reported on the plot, of $m_{Z'}$. The colored regions correspond to the exclusion by DD experiments.{\it Solid lines: Isocontours of the correct DM relic density for the same model but assuming the DM to be feebly interacting and produced via standard freeze-in.}}}
    \label{fig:pFIZp}
\end{figure}

$s$-channel portals are widely known to have a strong correlation between DM relic density and direct detection. Just for illustration, we show, in the case of the spin-1 portal (a similar outcome would occur in the other portal models), isocontours (dashed lines) corresponding to the correct DM relic density, in the $(m_\psi,g_\psi^V)$ bidimensional plane, for some fixed values of the mass of the mediator for $g_f^V=1$. The regions marked with the same color of the contours correspond to the current exclusion by Direct Detection experiments from the same assignations of the parameters. As evident, the conventional thermal paradigm is in very strong tension with experimental evidences. The freeze-in paradigm straightforwardly overcomes the problem as it allows to comply with the correct DM for very small values of the couplings. As evidenced by the solid contours of fig. \ref{fig:pFIZp} the parameter space corresponding to the correct relic density is very far from the experimental sensitivity.

As already pointed out, the $t$-channel portal model is less subject to constraints from Direct Detection, given the radiative origin of the SI interactions. Nevertheless, the conventional freeze-out might be disfavored given the $p$-wave suppression of the DM annihilation cross-section and the potential lower bounds on the mediator mass from LHC constraints \cite{Arina:2020tuw,Arina:2023msd}. Besides the conventional freeze-in, the $t$-channel mediators allows for other solutions for the the relic density ranging from coannihilations to conversion-driven freeze-out \cite{Garny:2017rxs,DAgnolo:2017dbv} and SuperWimp production \cite{Covi:1999ty,Feng:2003uy}.

In the case of electroweakly interacting DM the relic density depends only on the DM mass as annihilation processes occur via gauge interactions. The absence of tree level interactions between the DM and quark/gluons put this kind of models away from present DM sensitivity but in the reach of next generation detectors. However, the efficiency of DM annihilation allow to match the correct relic density only for heavy DM masses, starting from around $1\,\mbox{TeV}$ for a SU(2) doublet and increasing with the value of $n$. It would be then interesting to find lower DM mass benchmarks to be probed via DD experiments \footnote{Electroweakly interacting DM can be also effectively probed by Indirect Detection. A detailed study thereof is however beyond the purposes of this work}.

\subsection{Capability of Direct Detection in probing thermal freeze-out}

The paradigm of thermal freeze-out offers a straightforwardly testable solution to the puzzle of DM abundance generation in the Early Universe. However, the increasing sensitivity of direct detection (DD) experiments is placing increasingly stringent constraints on the magnitude of interactions between DM and SM states. This raises a fundamental question: would negative results at future DD facilities allow us to completely rule out the thermal paradigm, thereby indicating the necessity of alternative production mechanisms to solve the DM puzzle?

We first consider the simplified $s$-channel model with scalar DM. When the DM relic density is primarily determined by pair annihilations into SM fermion pairs, the thermally averaged cross-section can be reliably approximated using the so-called velocity expansion\footnote{We do not consider, in our study, values of the mass close to $s$-channel resonances or opening thresholds of annihilation channels, corresponding to the conventional cases where the velocity expansion is not valid \cite{Griest:1990kh}.}, which gives, at leading order:

\begin{align}
&   \langle \sigma v \rangle (\chi \chi \rightarrow  f f)
\approx \sum _f \frac{3 n_c^f}{16 \pi} {(\lambda_\chi^S)}^2 c_S^2 
\frac{m_f^2}{v_h^2}\frac{m_\chi^2}{(4 m_\chi^2-m_S^2)^2}\sqrt{1-\frac{4 m_f^2}{m_\chi^2}}\theta\left(m_\chi-m_f\right)\nonumber\\ 
& =  \frac{3}{16\pi}\frac{m_f^2}{v_h^2} {(\lambda_\chi^S)}^2 c_S^2 \sqrt{1-\frac{4 m_f^2}{m_\chi^2}}\theta\left(m_\chi-m_f\right) \times \left \{
\begin{array}{cc}
    \frac{m_\chi^2}{m_S^4} & m_S \gg m_\chi \\
    \frac{1}{16 m_\chi^2}  & m_S \ll m_\chi  
\end{array}
\right.
\end{align}

Using the velocity expansion, an analytical estimate of the relic density is given by \cite{Gondolo:1990dk}:
\begin{equation}
    \Omega_\chi h^2 \simeq 1.07\times 10^{9}{\mbox{GeV}}^{-1}\frac{x_f}{g_{*}^{1/2}M_{\rm Pl}\left(a+3b/x_f\right)},
\end{equation}
where $a$ and $b$ represent the $s$-wave and $p$-wave terms in the velocity expansion, respectively. Assuming $m_\chi \ll m_S$ and annihilations predominantly into bottom quark pairs, we can write:
\begin{equation}
    \left(\frac{\Omega_\chi h^2}{0.12}\right)\approx 19 {\left(\lambda_\chi^2 c_S^2\right)}^{-1} \left(\frac{m_\chi}{100\,\mbox{GeV}}\right)^2 \left(\frac{1/5}{m_\chi/m_S}\right)^4
\end{equation}

This equation can be solved for $\lambda_\chi^S c_S$ and the result substituted into the analytical expression for the DM scattering cross-section with protons:
\begin{equation}
    \sigma_{\chi p}^{\rm SI}= \frac{\mu_{\chi p}^2}{\pi}\frac{m_p^2}{m_\chi^2 m_S^4} (\lambda_\chi^S)^2 c_S^2 f_p^2
\end{equation}
where $f_p=\sum_{q=u,d,s}f_q^p+\frac{6}{27}f_{TG}\approx 0.3$. In this way, it is possible to obtain the following prediction for the DM scattering cross-section as a function of its mass:
\begin{equation}
    \sigma_{\chi p}^{\rm SI}\approx 1.1\times 10^{-42}{\left(\frac{100\,\mbox{GeV}}{m_\chi}\right)}^4 {\left(\frac{0.12}{\Omega_\chi h^2}\right)}^{-1}
\end{equation}

Repeating the same procedure for the other $s$-channel portals, we have, in the case of the fermionic DM with spin-0 mediator:

\begin{align}
    & \langle \sigma v \rangle(\bar \psi \psi \rightarrow  \bar f f)~
\approx \sum_f \frac{3 n_c^f v^2}{4\pi}g_\psi^2 c_S^2 \frac{m_f^2}{v_h^2}\frac{m_\psi^2}{(4 m_\psi^2-m_S^2)^2}\sqrt{1-\frac{4 m_f^2}{m_\chi^2}}\theta\left(m_\psi-m_f\right)\nonumber\\
& = \frac{3 n_c^f v^2}{4\pi}g_\psi^2 c_S^2 \frac{m_f^2}{v_h^2}\sqrt{1-\frac{4 m_f^2}{m_\chi^2}}\theta\left(m_\psi-m_f\right) \times \left \{
\begin{array}{cc}
    \frac{m_\psi^2}{m_S^4} & m_S \gg m_\psi \\
    \frac{1}{16 m_\psi^2}  & m_S \ll m_\psi  
\end{array}
\right.
\end{align}

For the vector portal, we have:
\begin{align}
    &  \langle \sigma v \rangle (\bar \psi \psi \rightarrow \bar f f)=\frac{\left \vert g_\psi^V \right \vert^2 }{\pi} \frac{m_\psi^2}{(4 m_\psi^2-m_{Z'}^2)^2}\sum_f n_f^c |g_f^{V}|^2 \sqrt{1-\frac{4m_f^2}{m_\psi^2}}\theta\left(m_\psi-m_f\right)\nonumber\\
    &= \frac{\left \vert g_\psi^V \right \vert^2}{\pi} \sum_f n_f^c|g_f^{V}|^2 \sqrt{1-\frac{4m_f^2}{m_\psi^2}}\theta\left(m_\psi-m_f\right) \times \left \{
\begin{array}{cc}
    \frac{m_\psi^2}{m_{Z'}^4} & m_{Z'} \gg m_\psi \\
    \frac{1}{16 m_\psi^2}  & m_{Z'} \ll m_\psi  
\end{array}
\right.
\end{align}
and 
\begin{equation}
    \sigma_{\psi p}^{\rm SI}=\frac{\mu_{\psi p}^2}{\pi}\frac{\left \vert g_\psi^V \right \vert^2 \left \vert 2 g_u^V+ g_d^V \right \vert ^2}{m_{Z'}^4}
\end{equation}
for the DM annihilation and scattering cross-sections, respectively. Considering the hierarchy $m_\psi \ll m_{Z'}$, we can translate the requirement of correct relic density into a relatively simple equation:
\begin{equation}
    \left(\frac{\Omega_\psi h^2}{0.12}\right) \approx 3.5 \times 10^{-2} \left(\sum_f n_c^f \left \vert g_\psi^V \right \vert^2 \left \vert g_f^V \right \vert^2\right)^{-1} \left(\frac{m_\psi}{100\,\mbox{GeV}}\right)^2 \left(\frac{1/5}{m_\psi/m_{Z'}}\right)^4 
\end{equation}
whose result can be related to the following prediction for the scattering cross-section with protons:
\begin{equation}
    \sigma_{\psi p}^{\rm SI} \approx 2.3 \times 10^{-40}\,\mbox{cm}^2 \frac{\left \vert 2 g_u^V+g_d^V \right \vert^2}{\sum_f n_f^c |g_f^V|^2}{\left(\frac{100\,\mbox{GeV}}{m_\psi}\right)}^2 {\left(\frac{0.12}{\Omega_\psi h^2}\right)}
\end{equation}

\begin{figure}
    \centering
    \includegraphics[width=0.5\linewidth]{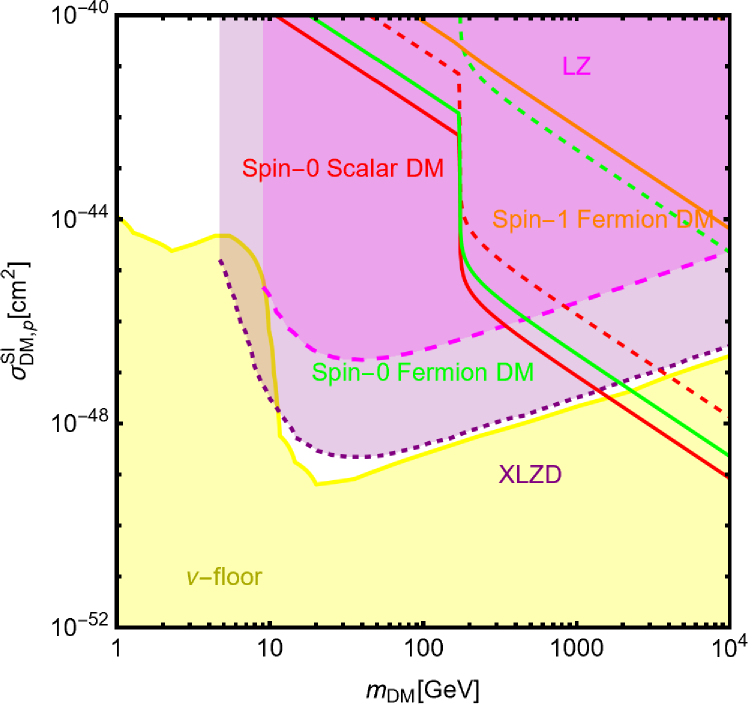}
    \caption{\footnotesize{Expected scattering cross-section over proton for some simplified setups. For each line the ratio between the DM mass and the mass of the mediator is kept fixed; solid lines refer to the case $M_{\rm med}=5 M_{\rm DM}$ while dashed lines correspond to the case $M_{\rm med}=1/5 M_{\rm DM}$. The couplings have been fixed in order to reproduce the correct DM relic density according to the thermal freeze-out paradigm.}}
    \label{fig:ptotsimp}
\end{figure}

The analytical results just discussed are illustrated comprehensively in Fig.~\ref{fig:ptotsimp}. The figure shows the $(m_{\rm DM},\sigma_{\rm DM p}^{\rm SI})$ parameter space, where the solid lines correspond to the expected DM scattering cross-section (with different colors corresponding to different portal models as displayed directly on the plot) when the DM couplings are fixed to achieve the correct relic density, assuming $m_{\chi,\psi}=1/5 m_{S,Z^{'}}$. The dashed lines have been obtained in an analogous fashion but considering $m_{\chi,\psi}=5 m_{S,Z^{'}}$.

The results clearly demonstrate that when a strong correlation between DM relic density via annihilation into SM quarks and DM scattering with nucleons is enforced, the freeze-out paradigm is already in strong tension with experimental data. An absence of detection at future facilities would seemingly exclude DM coupled with quarks below the TeV scale.

A possible way out, even within simplified setups such as those under consideration, would be to consider scenarios where the DM relic density relies essentially on annihilations into pairs of mediators (this implicitly requires heavier DM relative to the $s$-channel mediator). In this way, one can achieve effective DM annihilations without necessarily implying a scattering cross-section in tension with experimental data (for a detailed study of this scenario, the reader may refer to \cite{Arcadi:2016qoz}).

In this ``secluded'' regime, assuming the hierarchy $m_{\chi,\psi} \gg m_{S,Z^{'}}$ so that final state masses can be neglected, the relic density depends only on the DM mass and a single coupling. Considering the following cross-sections:

\begin{align}\label{eq:dmpairtoss}
    & \langle \sigma v \rangle (\chi \chi \rightarrow S S) \approx \frac{{(\lambda_\chi^S)}^4}{64 \pi m_\chi^2},\nonumber\\
   & \langle \sigma v \rangle(\bar \psi \psi \rightarrow S S) \approx 
 \frac{3}{64\pi}g_\psi^4 \frac{1}{m_\psi^2}v^2,\nonumber\\
 & \langle \sigma v \rangle (\bar \psi \psi \rightarrow Z' Z')=\frac{ |g_\psi^{V}|^4}{16\pi m_\psi^2},
\end{align}
we obtain the following predictions:
\begin{align}
\label{eq:lambda_sec}
    & \lambda_{\chi}^S \approx 0.24\, {\left(\frac{m_\chi}{100\,\mbox{GeV}}\right)}^{1/2} {\left(\frac{0.12}{\Omega_\chi h^2}\right)}^{1/4}\nonumber\\
    & g_\psi \approx 0.50\,{\left(\frac{m_\psi}{100\,\mbox{GeV}}\right)}^{1/2}{\left(\frac{0.12}{\Omega_\psi h^2}\right)}^{1/4}\nonumber\\
    & g_\psi^V \approx 0.22 \,{\left(\frac{m_\psi}{100\,\mbox{GeV}}\right)}^{1/2}{\left(\frac{0.12}{\Omega_\psi h^2}\right)}^{1/4}
\end{align}

Although it does not directly impact the relic density, the coupling between the mediator and the SM is subject to the requirement of ensuring DM thermalization in the Early Universe. The minimal value ensuring this condition can be estimated via the following condition:
\begin{equation}
    \left. \frac{\langle \sigma v \rangle (\rm DM \,DM \rightarrow \bar f f) n_{\rm DM,eq}}{H} \right \vert_{T=T_{\rm s.f.o.}}=1
\end{equation}

\begin{figure}
    \centering
    \subfloat{\includegraphics[width=0.45\linewidth]{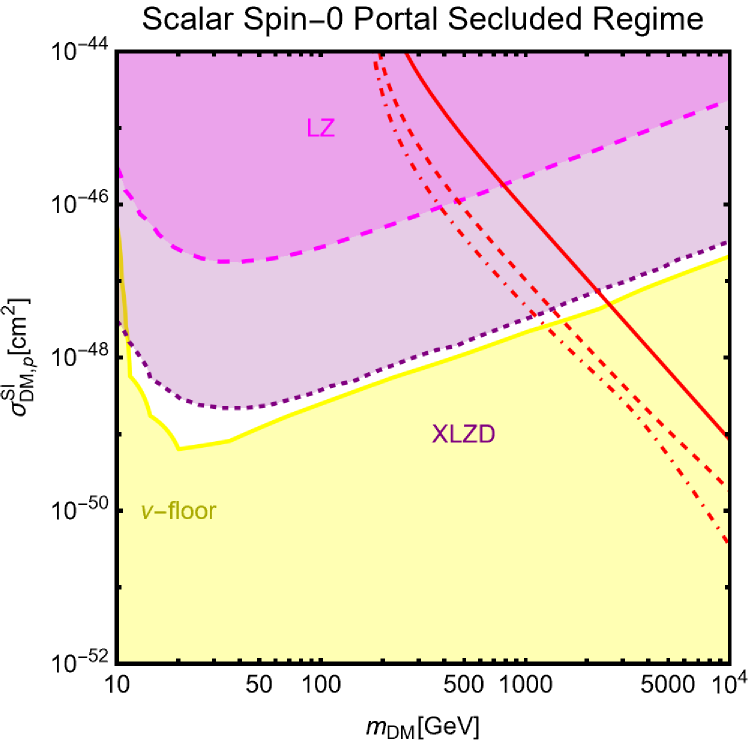}}
    \subfloat{\includegraphics[width=0.45\linewidth]{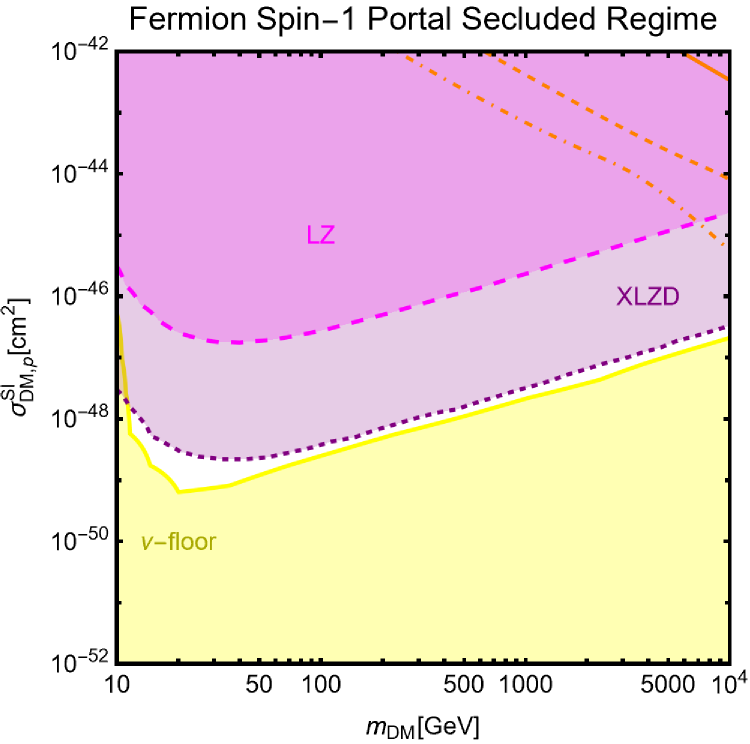}}
    \caption{\footnotesize{DM scattering cross-section over proton, as function of the DM mass, in simplified setups, indicated on top of the plots. The coupling between the DM and the mediator has been set in order to reproduce the correct relic density assuming that the latter is dominated essentially by $\mbox{DM} \mbox{DM} \rightarrow \mbox{Med}\mbox{Med}$ annihilations. The coupling between the mediator and the SM states has been set to the minimal value ensuring that the DM was in thermal equilibrium in the Early Universe.}}
    \label{fig:ptotsec}
\end{figure}

The two panels of Fig.~\ref{fig:ptotsec} show contours in the $(m_{\rm DM},\sigma_{\rm DM p}^{\rm SI})$ parameter space. These correspond to the prediction of the DM scattering cross-section for the three $s$-channel portals under scrutiny, with coupling values according to Eqs.~(\ref{eq:dmpairtoss}) and (\ref{eq:lambda_sec}). The different lines in each plot correspond to different mass ratios: $m_{\chi,\psi}=3 m_{S,Z'}$ (solid lines), $m_{\chi,\psi}=5 m_{S,Z'}$ (dashed lines), and $m_{\chi,\psi}=10 m_{S,Z'}$ (dot-dashed lines).

Beyond the ``secluded regime,'' other possibilities allow for a less tight connection between relic density and direct detection. Indeed, one can have scenarios in which the effective coupling between the DM and nucleons is suppressed. Such suppression can be ``accidental'' as due to the so called ``blind-spots'' (in such a case the inclusion of effects emerging at one or more loop levels is crucial to assess the capability of DD experiments to test this class of models) or more intrinsic of the model under scrutiny. A relevant example of this last kind of scenario is represented by models in which the DD is due to operators emerging at the one-loop level.

First, we can have suppression of the effective couplings between DM and nucleons in the presence of so-called blind spots. However, blind spots normally correspond to very specific assignments of model parameters and are not stable under radiative corrections. The inclusion of effects emerging at one or more loop levels is crucial to assess the capability of DD experiments to test this class of models.
The corresponding cross-sections have been computed in detail, for example, in \cite{Hisano:2011cs} and \cite{Arcadi:2023imv}, for electroweakly interacting and $t$-channel mediator DM, respectively. Another relevant example is represented by DM interacting with $s$-channel simplified mediators coupled to leptons rather than quarks. We will discuss this type of scenario later in the paper.
The capability of next generation experiments of probing concrete models with suppressed DD has been recently discussed in \cite{Arcadi:2025sxc}.

Another interesting possibility would be represented by the case that the thermally produced DM represents just a fraction of the total DM component of the Universe. This kind of scenario will be object of future work.

The case of study of present paper is instead represented by single component DM produced, possibly non thermally, in non-standard cosmological conditions. The correct relic density will be in general achieved in different regions of the parameter space with respect to the conventional freeze-out paradigm. One of the main purposes of present study will be to assess whether at least some of such regions will be accessible to next future experiments.


\section{DM production in Non-Standard Cosmologies: }

The standard thermal freeze-out and freeze-in paradigms for dark matter (DM) production are predicated on the assumption that DM genesis occurs during a radiation-dominated epoch in the early Universe. However, this assumption may not hold in more general cosmological scenarios, where the expansion history could be significantly modified by the presence of additional energy components beyond the standard model radiation bath. Such deviations from the standard cosmological model can profoundly alter the DM abundance calculation and open new parameter space regions that would otherwise be inaccessible to experimental probes.

In this section, we explore DM production mechanisms in non-standard cosmological scenarios, focusing particularly on the case where an exotic energy component $\Phi$ dominates the Universe's energy budget during the epoch of DM production. This component can serve both as a modifier of the cosmological expansion rate and as a potential source of DM particles through its decay. The interplay between these effects leads to rich phenomenology that can significantly expand the viable parameter space for DM models, particularly in regions that may be probed by current and future direct detection experiments.

The departure from standard cosmology introduces several new parameters that characterize both the exotic component and its coupling to the DM sector. These include the equation of state parameter $\omega$ of the exotic component, its decay rate $\Gamma_\Phi$, and, finally, the parameter $b_{\rm DM}$, determining the proportion between decays into radiation and DM particles. More specifically we are implicitly assuming that the $\Phi$ field features one (or more) decay channels into only SM final states as well as a decay channel into a DM pair. In this setup $b_{\rm DM}$ is just twice the branching ratio of the $\Phi \rightarrow \mbox{DM}\,\mbox{DM}$ process. The resulting DM abundance depends sensitively on these parameters, as well as on the reheating temperature $T_R$ at which the Universe transitions back to radiation domination, which must occur before Big Bang Nucleosynthesis (BBN) to preserve the success of standard cosmology at later times.
The setup described above can be translated into the following set of Boltzmann's equations:

\begin{align}
\label{eq:Boltzmann}
    & \frac{d\rho_\phi}{dt}+3 (1+\omega)H \rho_\phi=-\Gamma_\phi \rho_\phi \nonumber\\
    & \frac{ds}{dt}+3Hs=\frac{\Gamma_\phi \rho_\phi}{T}\left(1-b_{\rm DM}\frac{E_{\rm DM}}{m_\phi}\right)+2\frac{E_{\rm DM}}{T}\langle \sigma v \rangle \left(n_{\rm DM}^2-n^2_{\rm DM,\rm eq}\right) \nonumber\\
    & \frac{dn_{\rm DM}}{dt}+3Hn_{\rm DM}=\frac{b_{\rm DM}}{m_\phi}\Gamma_\phi \rho_\phi-\langle \sigma v \rangle \left(n_{\rm DM}^2-n^2_{\rm DM, \rm eq}\right)
\end{align}
where $E_{\rm DM}=\sqrt{m_{\rm DM}^2+T^2}$.
A rather general study of the scenario with $b_{\rm DM}=0$, namely no DM production from $\Phi$, has been presented in \cite{Arias:2019uol} (see also \cite{Arcadi:2024jzv}). Here we focus on some representative scenarios where the exotic component can directly contribute to DM production.

\begin{figure}
    \centering
    \subfloat{\includegraphics[width=0.35\linewidth]{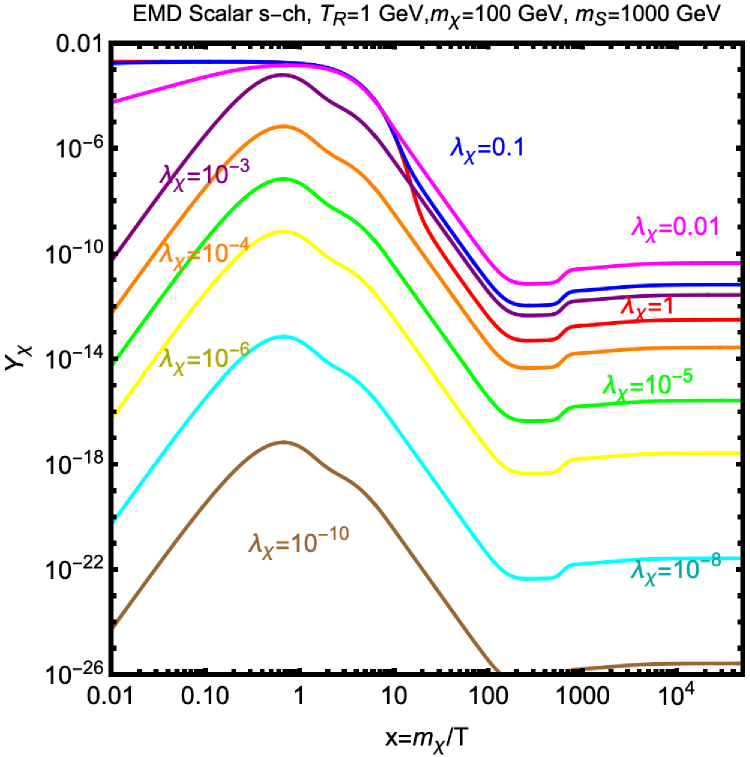}}
    \subfloat{\includegraphics[width=0.35\linewidth]{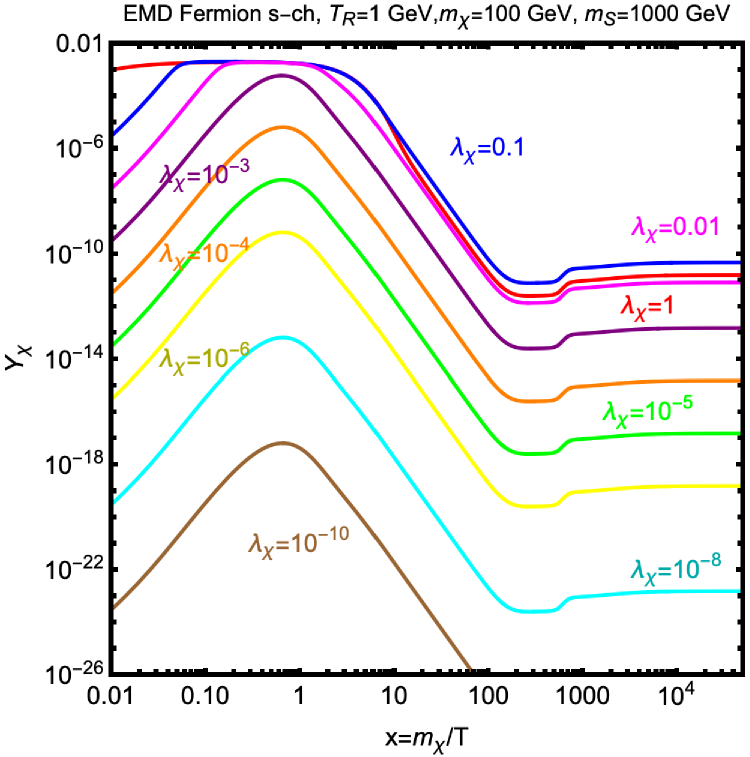}}
    \subfloat{\includegraphics[width=0.35\linewidth]{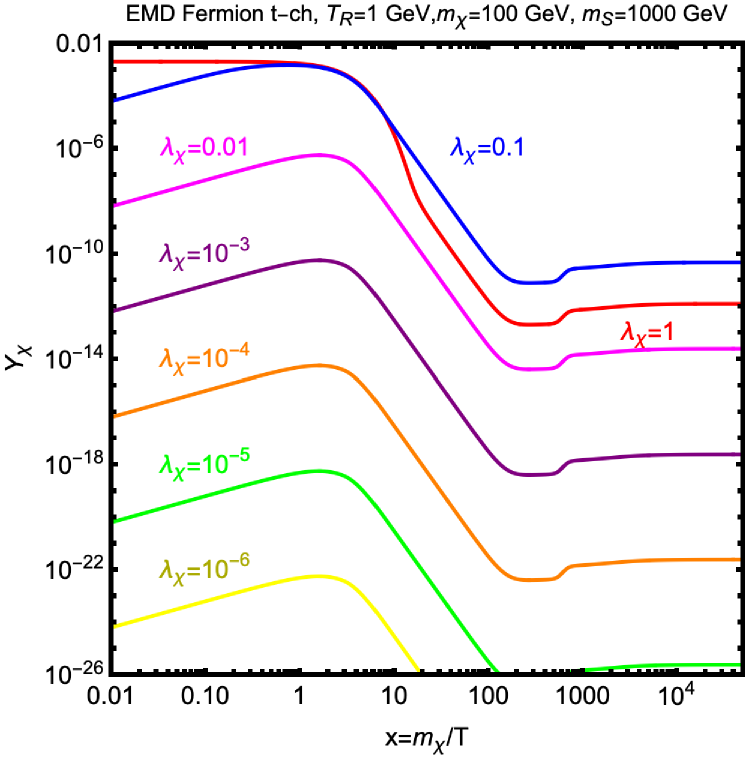}}
    \caption{\footnotesize{DM Yield in a scenario of thermal production with a non standard cosmology with Early Matter Domination (EMD). The three panels correspond to different simplified models, indicated at the top of each panel. In each plot the different colored lines correspond to different values of the DM coupling, reported in vicinity of each line.}}
    \label{fig:pEMDbol}
\end{figure}

\begin{figure}
    \centering
    \subfloat{\includegraphics[width=0.35\linewidth]{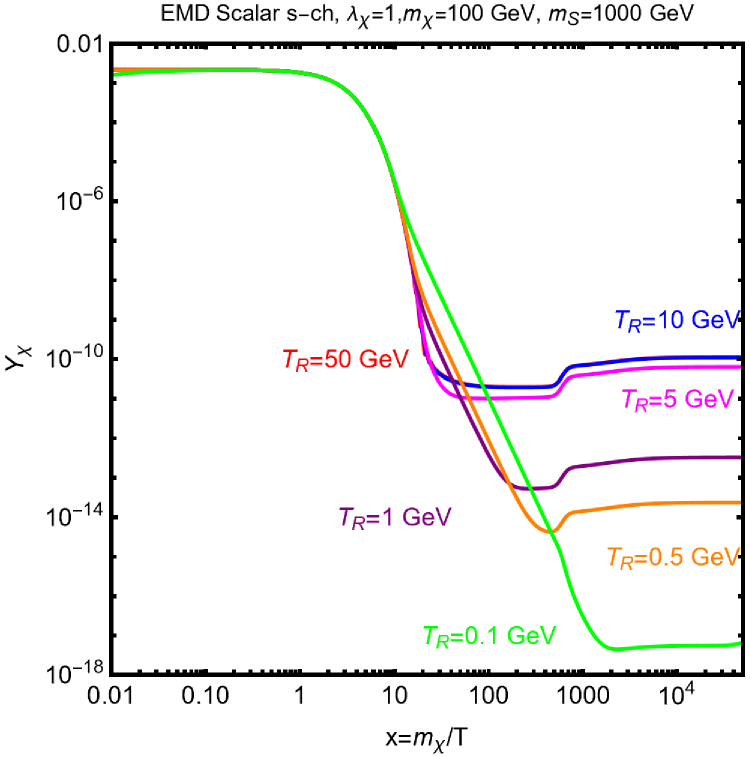}}
    \subfloat{\includegraphics[width=0.35\linewidth]{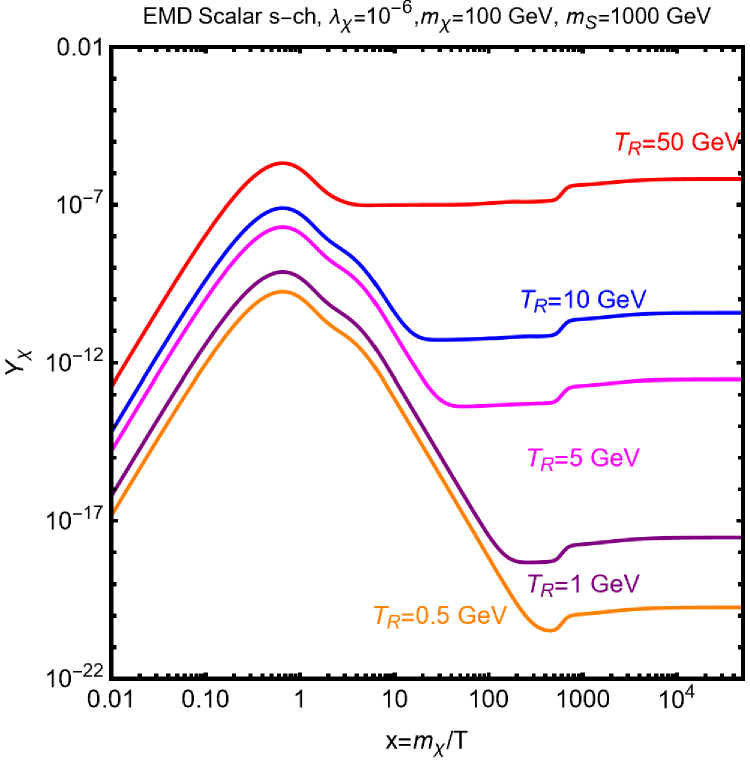}}
    \subfloat{\includegraphics[width=0.35\linewidth]{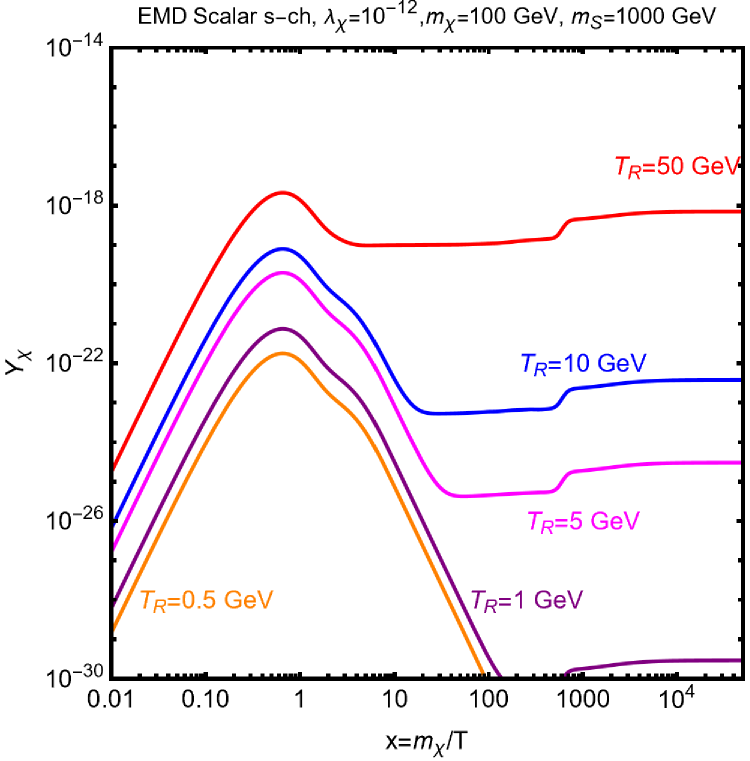}}\\
    \subfloat{\includegraphics[width=0.35\linewidth]{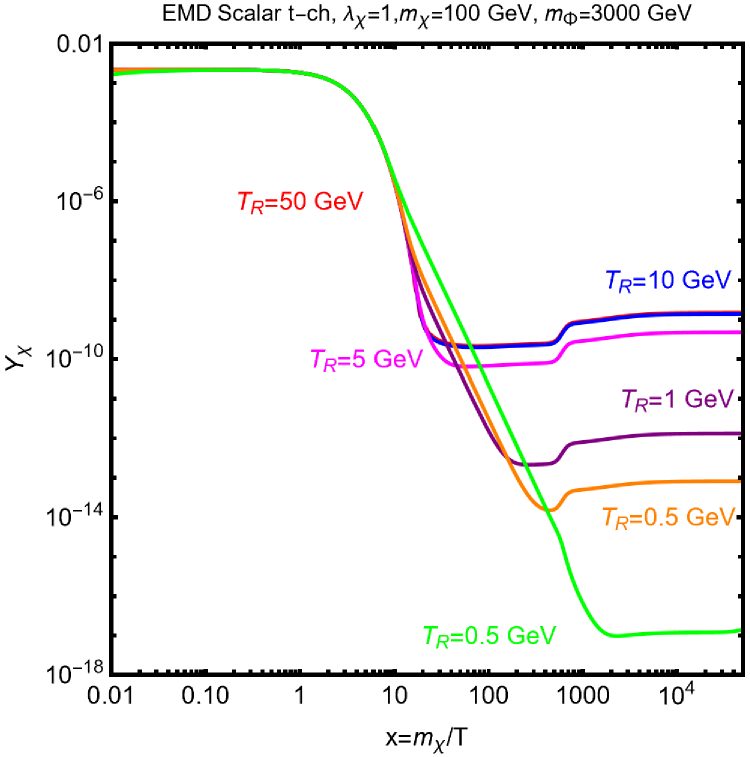}}
    \subfloat{\includegraphics[width=0.35\linewidth]{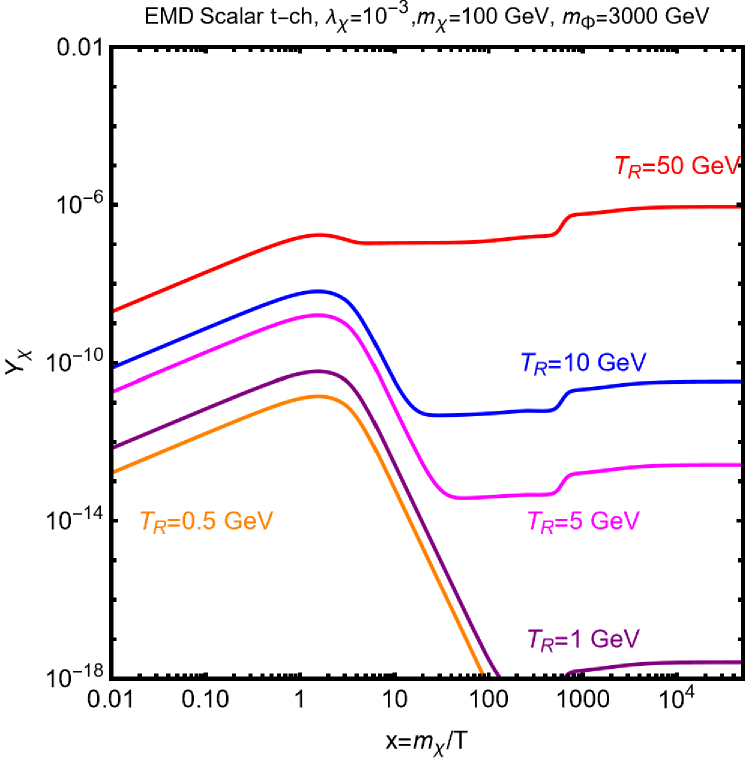}}
    \subfloat{\includegraphics[width=0.35\linewidth]{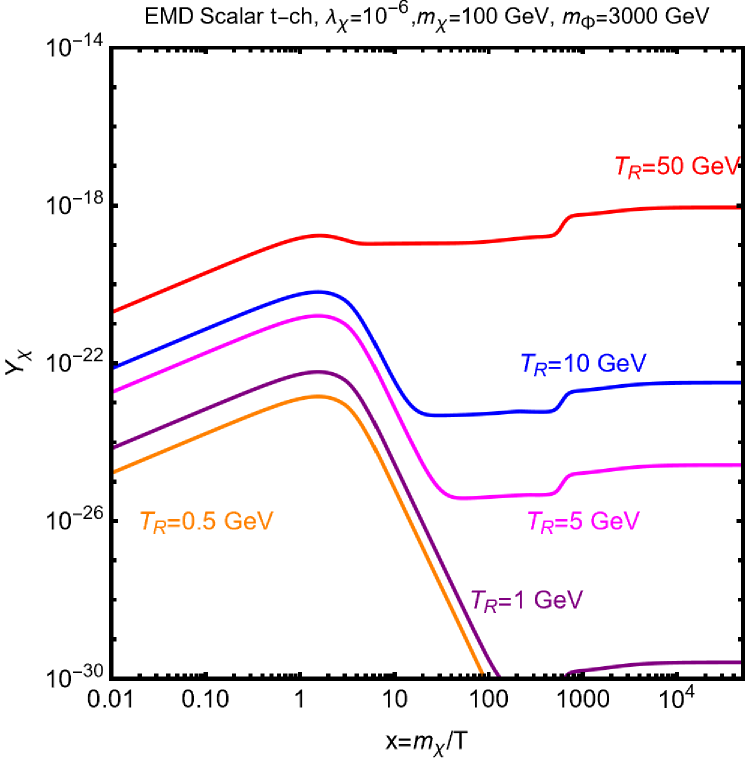}}
    \caption{\footnotesize{DM Yield, as function of the mass, in the scenario of DM production in a cosmological history with Early Matter domination. The first row refers to the $s$-channel simplified model with scalar DM while the $t$-channel simplified model has been, instead, considered, in the bottom row. In each panel the colored curves correspond to different values of the reheating temperature. In each row the different panels correspond to different values of the couplings, namely $\lambda_\chi=1,10^{-6},10^{-12}$ of the $s$-channel simplified model and $\lambda_\chi=1,10^{-3},10^{-6}$ for the $t$-channel simplified model. }}
    \label{fig:pEMDl}
\end{figure}

Notice that even if we are considering the field $\Phi$ as a generic additional component to the energy budget of the Universe, most of our findings can be applied as well to DM production during an inflationary (perturbative) reheating. For some more dedicated study the reader might refer, for example, to \cite{Bernal:2022wck,Silva-Malpartida:2023yks,Bernal:2024yhu,SilvaMalpartida:2024abc,Henrich:2025gsd}.

\subsection{Early Matter Domination Scenario}

We first consider the case customarily dubbed Early Matter Domination (EMD), corresponding to $\omega=0$. In this scenario, the exotic component $\Phi$ behaves as pressureless matter, leading to a modified expansion history where $H \propto a^{-3/2}$ rather than the standard radiation-dominated scaling $H \propto a^{-2}$. This altered expansion rate has profound implications for both the freeze-out dynamics and any non-thermal DM production processes.

The system of Boltzmann equations \eqref{eq:Boltzmann} can be solved analytically in this case via the following change of variables \cite{Giudice:2000ex,Arcadi:2011ev}:
\begin{equation}
    \Phi=\frac{\rho_\phi a^{3}}{\Lambda},\,\,\,\,\,N_{\rm DM}=n_{\rm DM} a^3,\,\,\,\,a=\frac{A}{a_I}
\end{equation}
This transformation allows us to factor out the terms linear in the Hubble expansion rate, recasting the system in the following more tractable form:

\begin{align}
\label{eq:NTDMbol}
    & \frac{d\Phi}{dA}=-\frac{\Gamma_\Phi}{\mathcal{H}}A^{1/2}a_I^{3/2}\Phi \nonumber\\
    & \frac{dN_{\rm DM}}{dA}=\frac{A^{1/2}a_I^{ 3/2}}{\mathcal{H}} \Lambda \frac{b_{\rm DM}}{m_\phi}\Gamma_\phi \Phi-\frac{\langle \sigma v \rangle}{\mathcal{H}}A^{-5/2}a_I^{-3/2}\left(N_{\rm DM}^2-N^2_{\rm DM,\rm eq}\right)\nonumber\\
    & \frac{dT}{dA}={\left(3+T\frac{dh_{\rm eff}}{dT}\right)}^{-1}\left \{- \frac{T}{A}+\frac{\Gamma_\phi \Lambda}{m_\phi}\left(1-\frac{b_{\rm DM} E_{\rm DM}}{m_\phi}\right)\frac{T}{s \mathcal{H}}A^{-5/2}a_I^{-3/2}\Phi \right. \nonumber\\
    & \left. +2 \frac{E_{\rm DM}}{s \mathcal{H}}A^{-11/2}a_I^{-9/2} \langle \sigma v \rangle \left(N_{\rm DM}^2-N^2_{\rm DM,\rm eq}\right) \right \}
\end{align}
where $\mathcal{H}$ is defined as:
\begin{equation}
    \mathcal{H}\equiv (a_I A)^{3/2}H={\left(\frac{\Lambda \Phi+\rho_R(T)A^3 a_I^3 +E_{\rm DM} N_{\rm DM}}{3 M_{Pl}^2}\right)}
\end{equation}
Unless differently stated, we will assume initial conditions (corresponding to $A=1$ in the new variables) consisting into a value of $10^6\,\mbox{GeV}$ for the temperature (or equivalently for the energy density) of radiation and to a value $\rho_\phi=\frac{1}{2}m_\phi^2 M_{\rm Pl}^2$, with $m_\phi=10^6 \,\mbox{GeV}$ for the initial energy density of $\Phi$\footnote{The chosen initial conditions imply a fixed initial ratio between the initial energy density of the radiation and of the exotic component $\Phi$. One could consider the latter as an additional free parameter, see e.g. \cite{Arias:2019abc} with potential impact to the DM relic density. We will leave this possibility to future study.}. The initial DM abundance is assumed to be zero.

Let's first consider the case $b_{\rm DM}=0$. A first illustration is provided by fig. \ref{fig:pEMDbol}. It shows, for the portal models, the evolution of the DM Yield, for a benchmark assignation of the DM and mediator masses and for the value of $1\,\mbox{GeV}$ of the reheating temperature. In each panel, the different lines correspond to the different value of the couplings between the DM and the mediator. By comparing the different contours one notices that for the highest values of the couplings the DM abundance can reach, at early stages the equilibrium curve. At later stages, instead of encountering the standard freeze-out, the DM abundance is depleted by the entropy injection due to the decay of $\Phi$. At the end of the decay process the DM comoving abundance reaches the final value corresponding to the relic density. In such a setup, the relic density appears to increase by lowering the value of the DM coupling. As the value of $\lambda_\chi$ decreases, the DM comoving abundance is not able to match at any time the equilibrium value and we notice a behavior more similar to the freeze-in, consisting into an initial progressive increase of the DM abundance, followed by the dilution from entropy injection in the plasma ending at around the reheating temperature. In this second regime the DM relic density is proportional to the DM coupling. The transition between the aformentioned freeze-out like and freeze-in like regimes depends on the models under scrutiny. In particular it occurs before in the $t$-channel simplified model due to the stronger dependence on the coupling $\lambda_\chi$ of the DM annihilation processes. In summary one can infer that, in a EMD one would need, to match the correct DM relic density, smaller couplings, with respect to the case of standard cosmological history, in the case of freeze-out and, instead, smaller couplings, always compared to the standard scenario, in the case of freeze-in. This would allow, in the former case, to evade the constraints from Direct Detection while, in the latter case, would allow the model under scrutiny to approach the experimental sensitivity, something normally very difficult in the standard freeze-in. 

Fig. \ref{fig:pEMDl} illustrates instead the behaviour of the DM relic density by varying the reheating temperature. For each model (the different rows in the figure) we have considered three representative values of the DM couplings, corresponding to the WIMP and FIMP regimes as well as to an intermediate regime among the two.

We are, finally, ready to assess the detection prospects of the scenarios of DM production during EMD. We have considered first the scenario of high values of the couplings, corresponding to thermal freeze-out in a standard cosmological scenario. We have performed a parameter scan over the following ranges: 
The scan ranges are:
\begin{align}
    & m_{\chi,\psi} \in \left[1,1000\right]\,\mbox{GeV},\,\,\,\,m_S \in \left[10\,\mbox{GeV},10\,\mbox{TeV}\right],\,\,\,\,\,\lambda_\chi^S, g_\psi \in \left[10^{-2},10\right]
\end{align}
for the two $s$-channel simplified models, 
\begin{align}
    & m_{\psi} \in \left[1,1000\right]\,\mbox{GeV},\,\,\,\,m_{Z'} \in \left[1,10\right]\,\mbox{TeV},\,\,\,\,\,g_\psi^V,g_f^V \in \left[10^{-2},10\right]
\end{align}
for the spin-1 portal, and
\begin{align}
     & m_\chi \in \left[1,1000\right]\,\mbox{GeV},\,\,\,\,m_\Phi \in \left[1,10\right]\,\mbox{TeV},\,\,\,\,\,\lambda_\chi \in \left[10^{-2},10\right]
\end{align}
for the $t$-channel simplified model. For all the models we have varied the reheating temperature in the range $\left[10\,\mbox{MeV},\min(10\,\mbox{GeV},T_{s.f.o.})\right]$.
\begin{figure}
    \centering
    \subfloat{\includegraphics[width=0.5\linewidth]{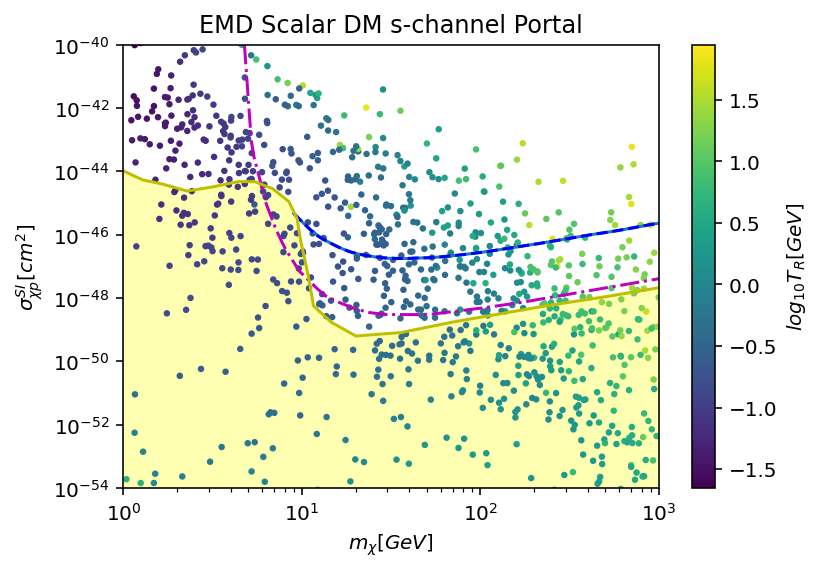}}
    \subfloat{\includegraphics[width=0.5\linewidth]{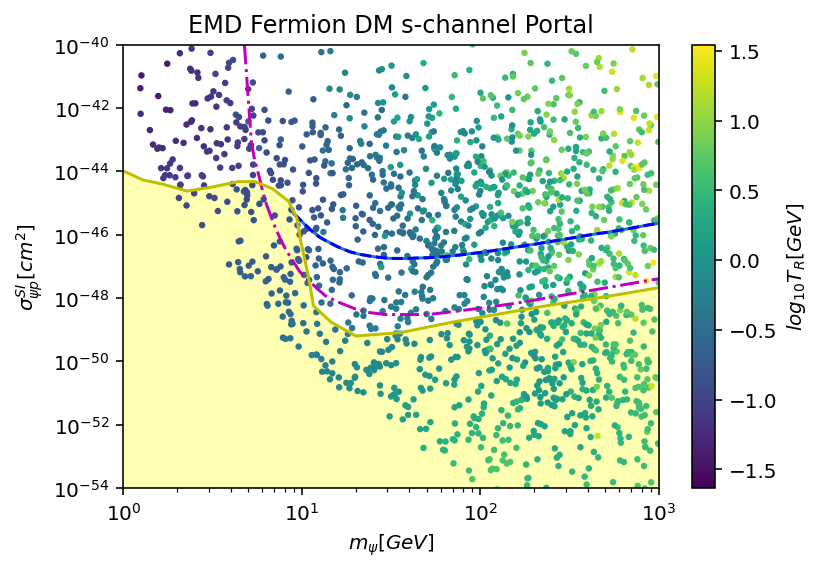}}\\
    \subfloat{\includegraphics[width=0.5\linewidth]{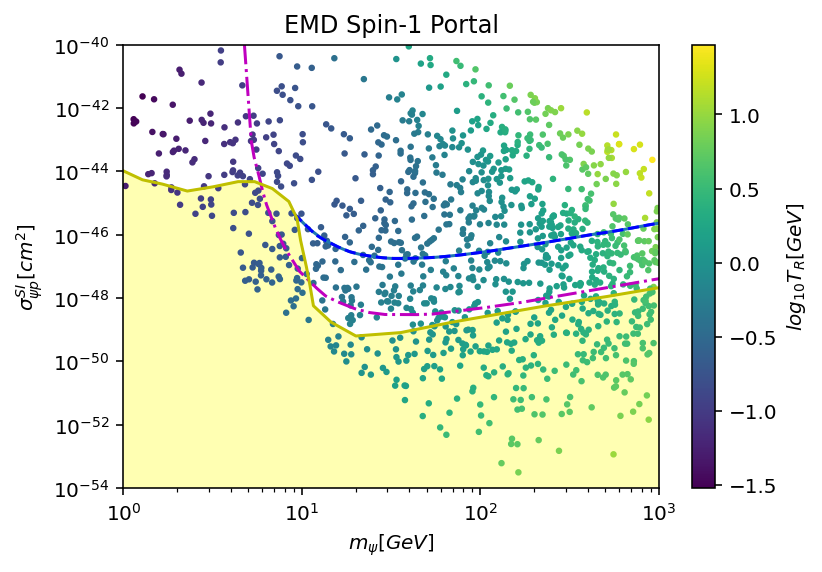}}
    \subfloat{\includegraphics[width=0.5\linewidth]{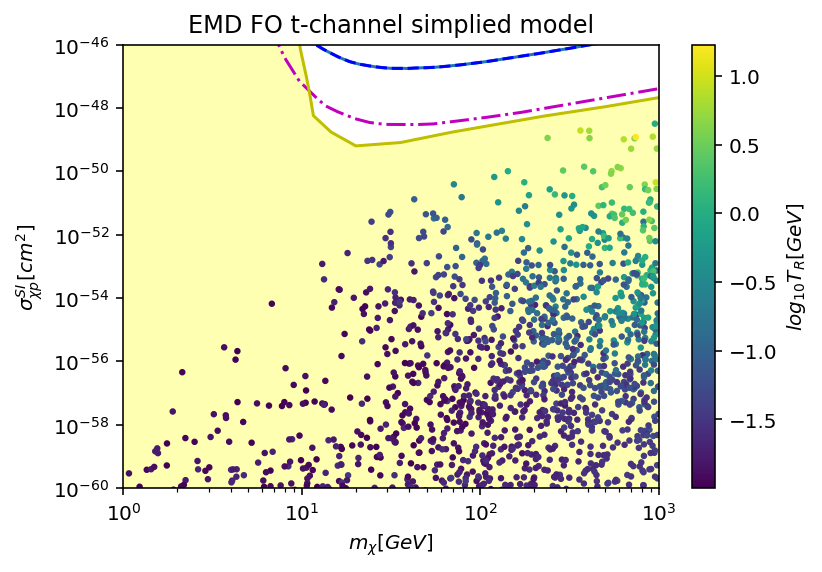}}
    \caption{Outcome of a parameter scan for the four portal simplified models assuming thermal production in a non-Standard Cosmological scenario with Early Matter domination.}
    \label{fig:pEDM}
\end{figure}

The results are shown in fig. (\ref{fig:pEDM}). The four panels (corresponding to the four portal models) of the figure show the model points featuring the correct relic density in the $(m_{\rm DM},\sigma_{DM,p}^{\rm SI})$ bidimensional plane. The current exclusion bound from LZ and the projected sensitivity by XLZD (assuming 200ty exposure) are reported as well as, respectively, dashed blue and dot-dashed purple lines. Finally, the region colored in yellow correspond to the neutrino floor. The color code of the model points correspond to the value of the reheating temperature.

\begin{figure}
    \centering
    \subfloat{\includegraphics[width=0.4\linewidth]{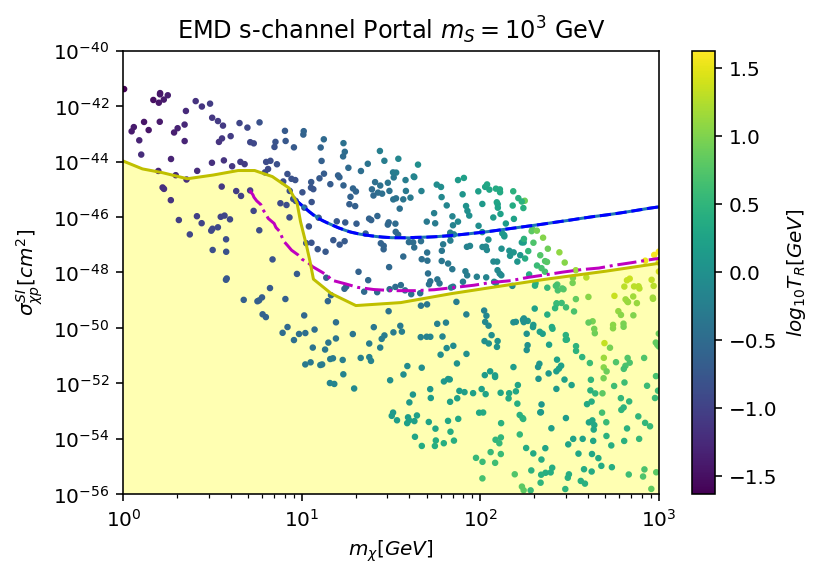}}
    \subfloat{\includegraphics[width=0.4\linewidth]{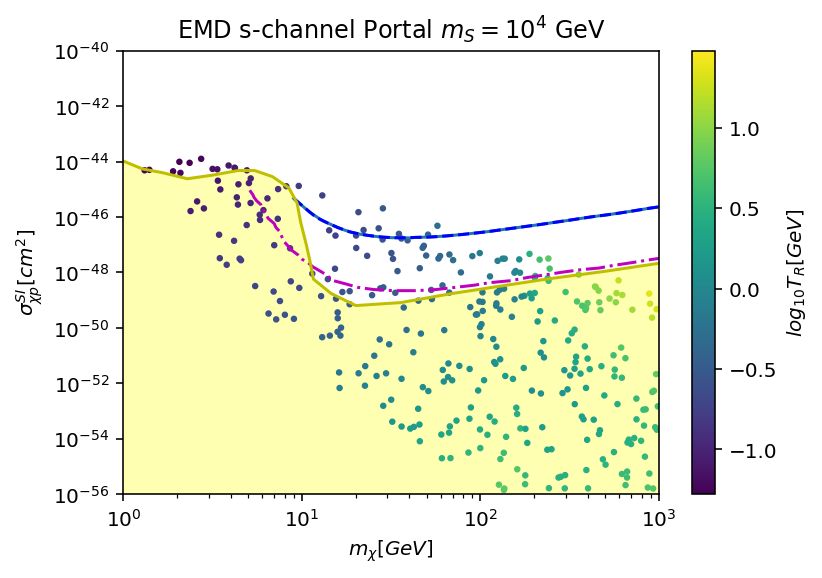}}
    \subfloat{\includegraphics[width=0.4\linewidth]{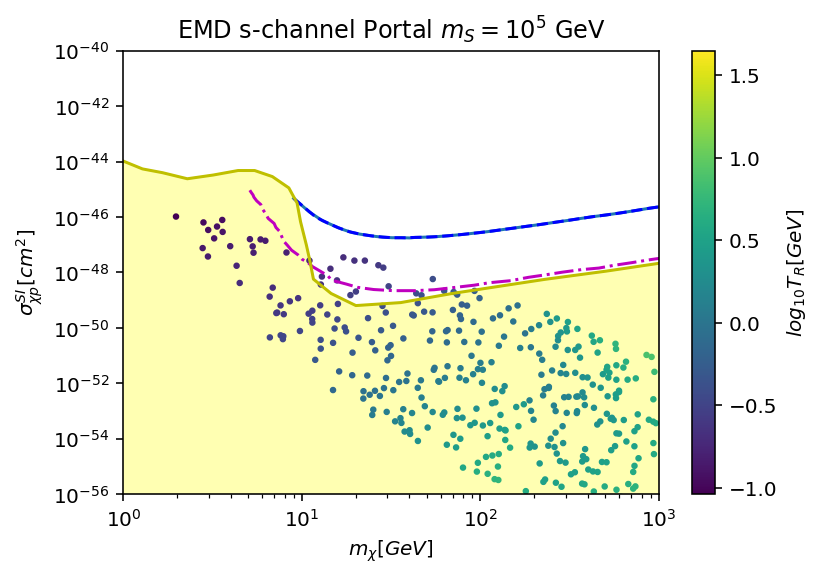}}\\
    \subfloat{\includegraphics[width=0.4\linewidth]{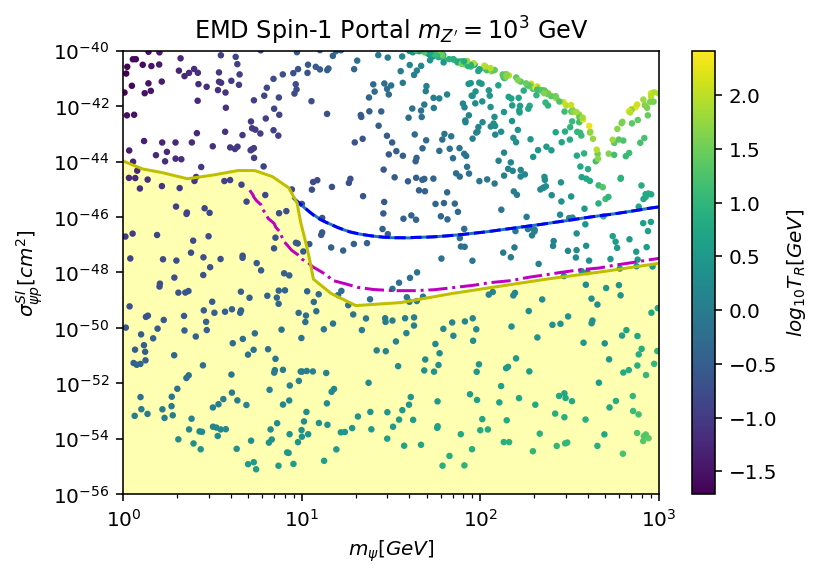}}
    \subfloat{\includegraphics[width=0.4\linewidth]{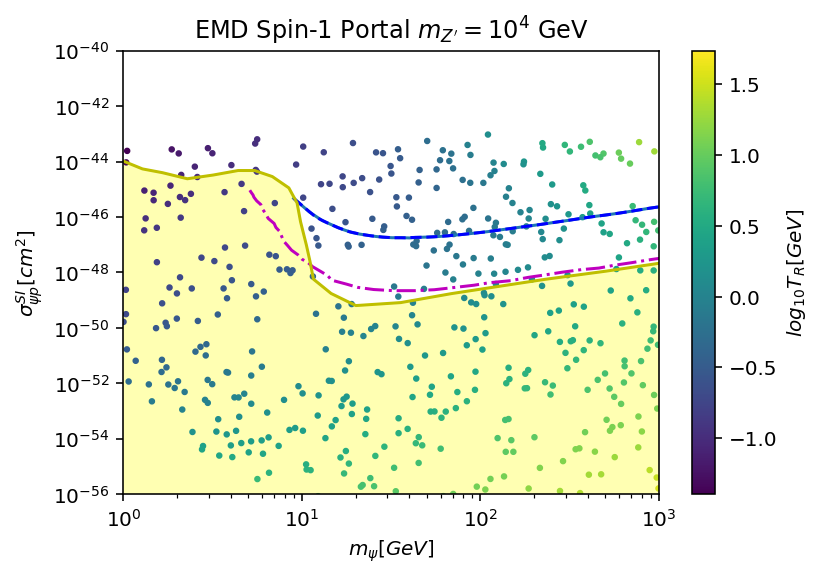}}
    \subfloat{\includegraphics[width=0.4\linewidth]{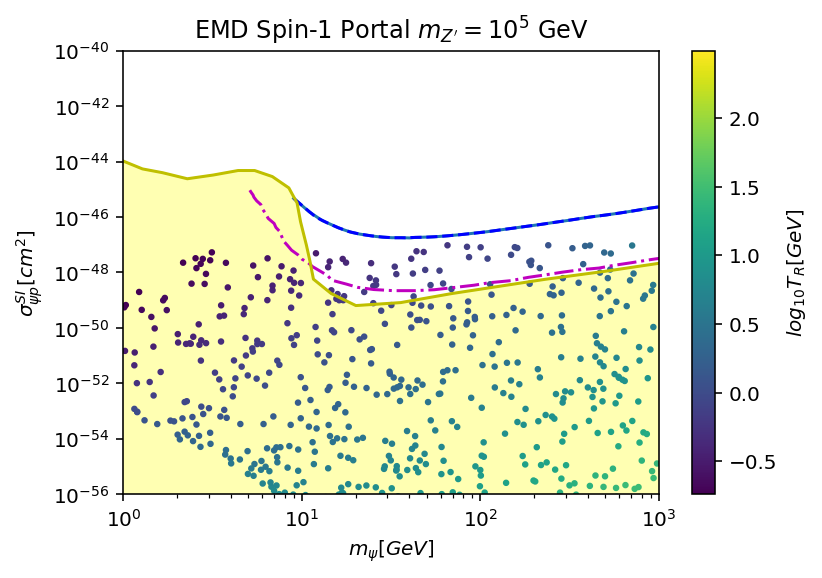}}
    \caption{\footnotesize{Outcome of parameter scan for the spin-1 portal, considering a wide of range of coupling values, to include both freeze-out and freeze-in regimes, (see main text for details). The three panels correspond to the three values $1,10,100\,\mbox{TeV}$ of the mass of the $Z'$ boson.}}
    \label{fig:pEMDhZp}
\end{figure}

Fig. \ref{fig:pEMDhZp} shows the results of a broader exploration of the parameter space in which a larger window of values of the DM coupling has been considered. For simplicity we have focused on just the $s$-channel portal with scalar DM and the spin--1 portal. In both cases the coupling has been varied in the $[10^{-10},1]$ range while the mass of the mediator has not been varied, contrary to the previous scan, but, instead, kept fixed to the three values of 1,10 and 100 TeV.

Up to now the simplified model with electroweak interacting DM has not been considered so far. The reason is that the DM annihilation cross-section depends on the SM gauge couplings and, for the considered range of masses, it leads to an underabundant density if the conventional thermal production is assumed. Consequently, the correct relic density can never be achieved in the case of a cosmology with an Early Matter domination, unless an extra non-thermal source of DM is assumed, as will be evidenced in the next section.


\subsection{Non-thermal DM production in EMD}

Let's know consider the case $b_{\rm DM} \neq 0$, i.e. the possibility of non-thermal DM production from the decay of the field $\Phi$.

\subsubsection{Reannihilation and Direct Production Regimes}

As in the previous subsection, we first illustrate the solution of the Boltzmann equations for some benchmarks.

\begin{figure}
    \centering
    \subfloat{\includegraphics[width=0.33\linewidth]{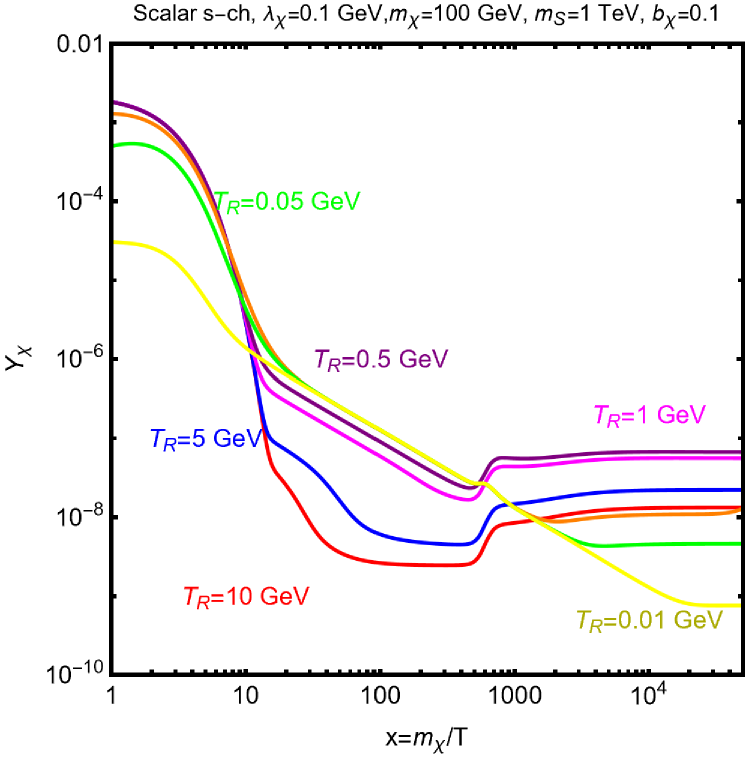}}
    \subfloat{\includegraphics[width=0.33\linewidth]{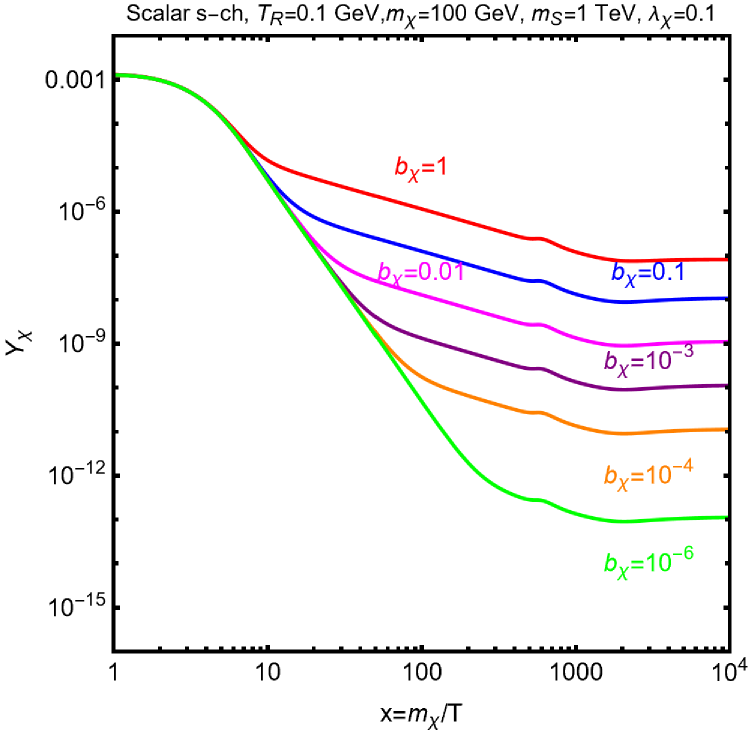}}
    \subfloat{\includegraphics[width=0.33\linewidth]{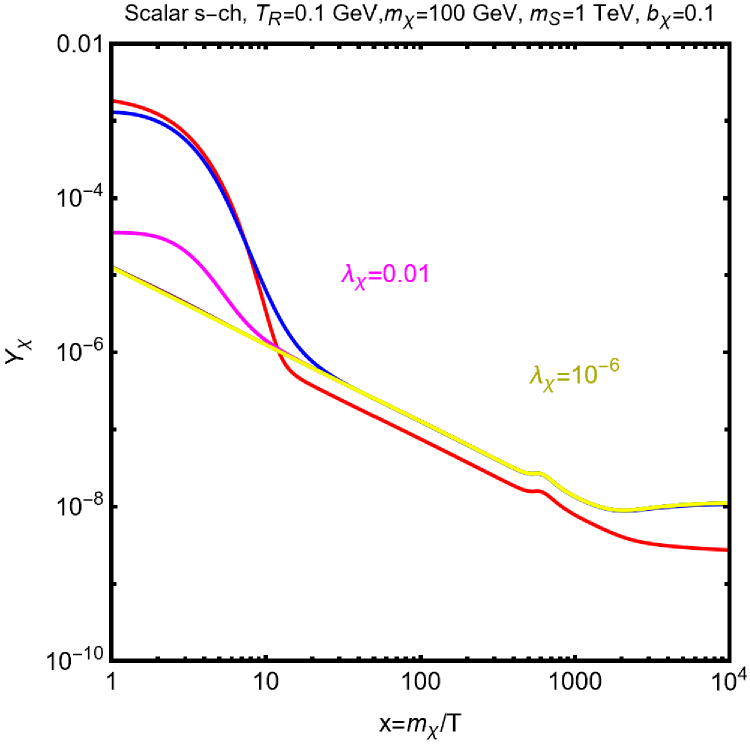}}\\
    \subfloat{\includegraphics[width=0.33\linewidth]{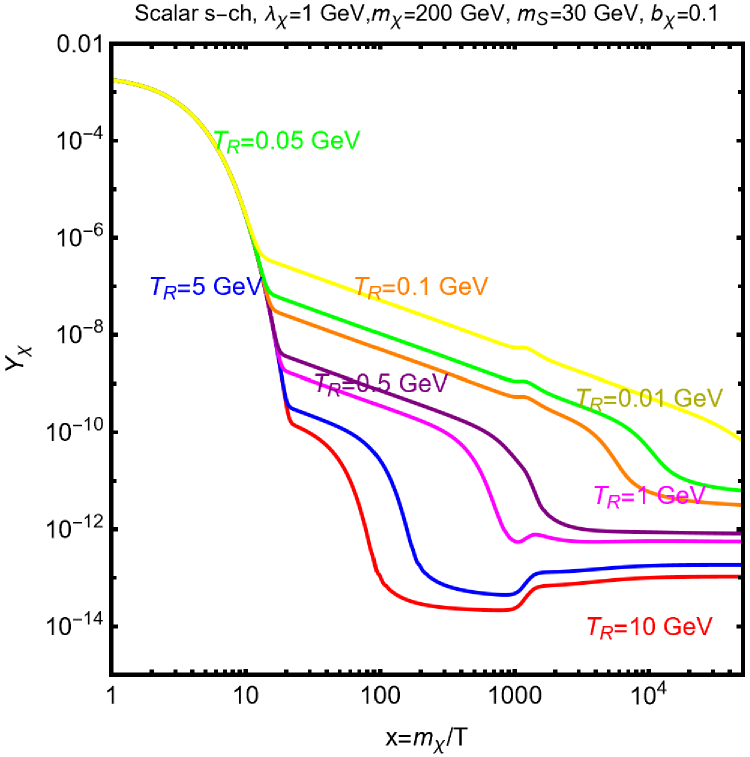}}
    \subfloat{\includegraphics[width=0.33\linewidth]{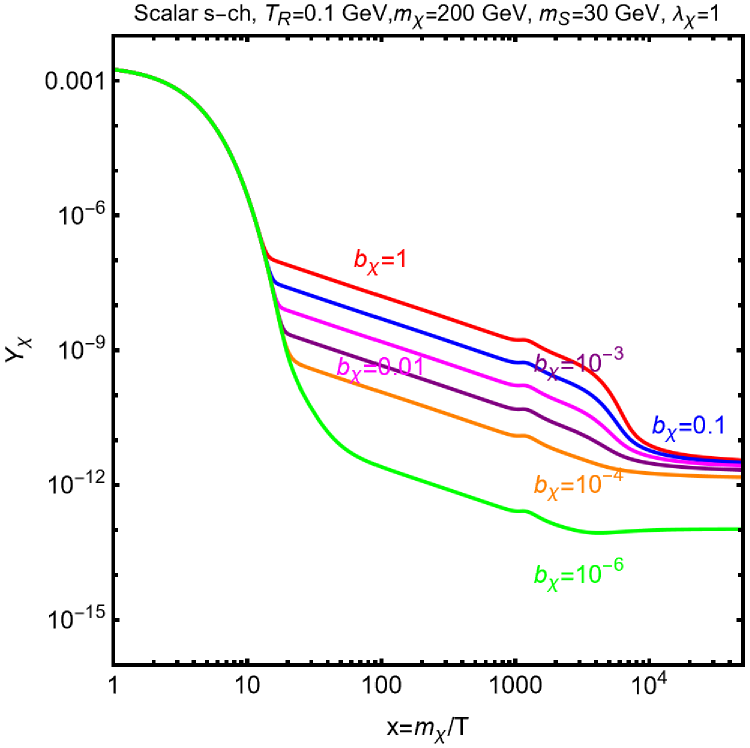}}
    \subfloat{\includegraphics[width=0.33\linewidth]{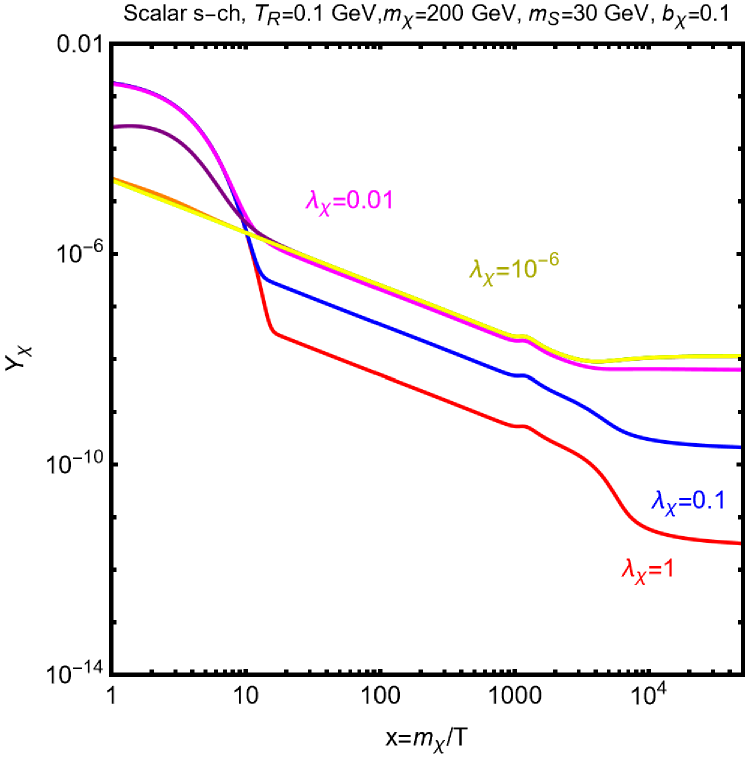}}
    \caption{\footnotesize{Evolution of the DM Yield as function of $x=m_\chi/T$ for the simplified $s$-channel model with scalar DM. The left column describes $Y_\chi$ for different values of the reheating temperature and fixed $b_\chi,\lambda_\chi$. The central column refers to different values of $b_\chi$ and fixed $\lambda_\chi,T_R$ while, finally, in the right column different values of $\lambda_\chi$ and fixed $b_\chi,T_R$ are considered. The two rows correspond to two benchmarks with different mass hierarchies, namely $m_\chi < m_S$ (top row) and $m_\chi > m_S$ (bottom row).}}    
    \label{fig:pNTDMbensch}
\end{figure}

Fig.~\ref{fig:pNTDMbensch} considers the case of the scalar $s$-channel simplified model with scalar DM. We have selected two benchmark parameter sets: $(m_\chi,m_S,c_S) = (100\,\mbox{GeV},1\,\mbox{TeV},1)$ and $(200\,\mbox{GeV},30\,\mbox{GeV},1)$, and studied the variation of the DM yield $Y_\chi=n_\chi/s$ as a function of the key parameters $\lambda_\chi$, $T_R$, and $b_\chi$.

The behavior of the DM yield curves can be understood in terms of two distinct asymptotic regimes that emerge from the solutions of system \eqref{eq:NTDMbol} (for more details see e.g. \cite{Arcadi:2011ev,Gelmini:2006pq}). The transition between these regimes depends critically on whether the DM annihilation rate is sufficient to deplete the non-thermally produced DM abundance.

When the DM annihilation rate is sufficiently large, the non-thermal abundance exceeds the critical value:
\begin{equation}
    n_{\rm DM}^c \simeq \frac{H}{\langle \sigma v \rangle}
\end{equation}
This triggers a reactivation of annihilation processes, leading to what is known as the ``reannihilation'' regime. This regime is realized when \cite{Aparicio:2016qqb}:
\begin{equation}
\label{eq:reannihilation}
    \langle \sigma v \rangle > 1.96 \times 10^{-29}\,\mbox{cm}^3 \mbox{s}^{-1}\, b_{\rm DM}^{-1}{\left(\frac{1 \,\mbox{GeV}}{T_R}\right)}^{4/3}  
\end{equation}
In the reannihilation regime, the DM relic density can be approximated as:
\begin{equation}
\Omega_{\rm DM}^{\rm NT}h^2 \simeq \frac{T_{s.f.o.}}{T_{\rm R}}\Omega_{\rm DM}^{\rm T}h^2 
\end{equation}
where $T_{s.f.o.}$ represents the standard freeze-out temperature (i.e., as in a radiation-dominated era) and $\Omega_{\rm DM}^{\rm T}$ represents the relic density computed in the conventional freeze-out paradigm. Since $T_R < T_{s.f.o.}$, the reannihilation regime is particularly well-suited for DM candidates that would be underabundant in the standard thermal freeze-out scenario. Notably, the relic density depends on the DM thermally averaged cross-section through $\Omega_\chi^{\rm T}$, but is essentially independent of the branching fraction $b_{\rm DM}$.

Conversely, when condition \eqref{eq:reannihilation} is not satisfied, the non-thermal production is insufficient to trigger efficient DM annihilation. In this case, the energy initially stored in the exotic component $\Phi$ is entirely converted into DM particles, leading to a relic density approximately given by: 
\begin{align}
\label{eq:second_regime}
    & Y_{\rm DM} (T_{\rm R})=\frac{n_{\rm DM} (T_{\rm R})}{s (T_{\rm R})}\simeq \frac{b_{\rm DM}}{m_\phi}\frac{\rho_\phi (T_{\rm R})}{s(T_{\rm R})}\simeq \frac{3}{4}\frac{b_{\rm DM}}{m_\phi}T_{\rm R}\nonumber\\
    & \rightarrow \Omega_{\rm DM}^{\rm NT}h^2 \simeq 0.2 \times 10^4 b_{\rm DM} \frac{10 \,\mbox{TeV}}{m_\phi}\frac{T_{\rm R}}{1 \,\mbox{MeV}}\frac{m_{\rm DM}}{100\,\mbox{GeV}}
\end{align}
In contrast to the reannihilation regime, this ``direct production'' regime yields a relic density that is independent of the annihilation cross-section and scales linearly with the reheating temperature, rather than inversely as in the reannihilation case.

The benchmark results shown in Fig.~\ref{fig:pNTDMbensch} illustrate these regime transitions clearly. The first row considers a mass hierarchy with $m_\chi \ll m_S$. For $\lambda_\chi^S = b_\chi = 0.1$, the relic density follows the reannihilation regime for $T_R \gtrsim 0.5\,\mbox{GeV}$, while for $T_R=0.1\,\mbox{GeV}$ we observe the linear dependence on $b_\chi$ characteristic of the direct production regime \eqref{eq:second_regime}. In the case $m_\chi \gg m_S$, the DM relic density is dominated by the reannihilation regime unless very small values of $b_\chi \lesssim 10^{-6}$ and/or small DM coupling $\lambda_\chi^S \lesssim 0.01$ are considered.

\subsubsection{Model-Specific Predictions}

The reannihilation regime typically enables the correct relic density for models characterized by DM annihilation cross-sections significantly above the thermally favored value of $\langle \sigma v \rangle \sim 3 \times 10^{-26}\,\mbox{cm}^3\mbox{s}^{-1}$. A prominent example is a DM candidate charged under $SU(2)_L$ with mass below 1 TeV, which would satisfy these requirements due to its enhanced annihilation cross-section. Conversely, models with $p$-wave suppressed cross-sections are generally expected not to reach the reannihilation regime. In such cases, the non-thermal DM production is highly efficient, necessitating a very suppressed branching fraction $b_{\rm DM}$ to achieve the correct relic density.

\begin{figure}
    \centering
    \subfloat{\includegraphics[width=0.5\linewidth]{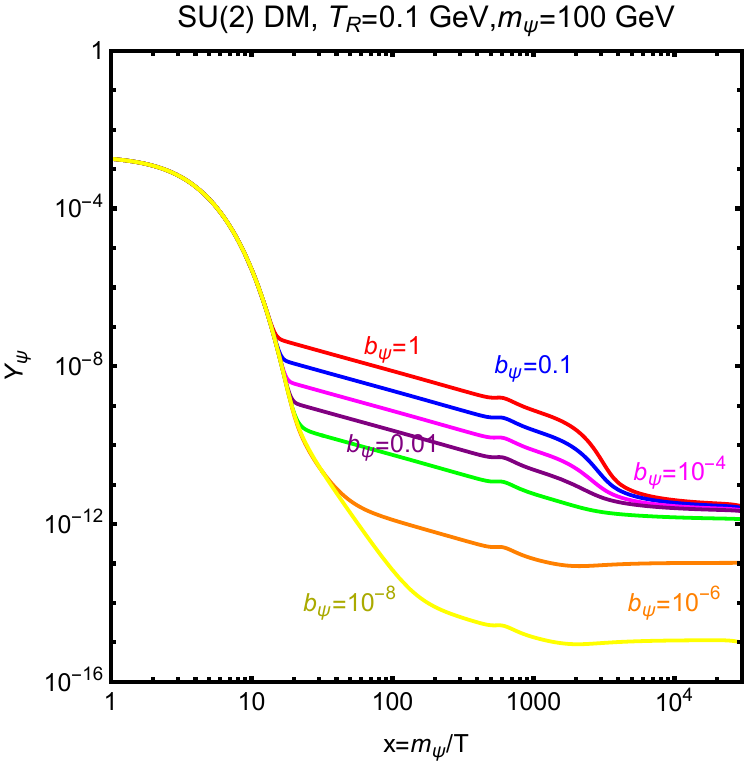}}
    \subfloat{\includegraphics[width=0.51\linewidth]{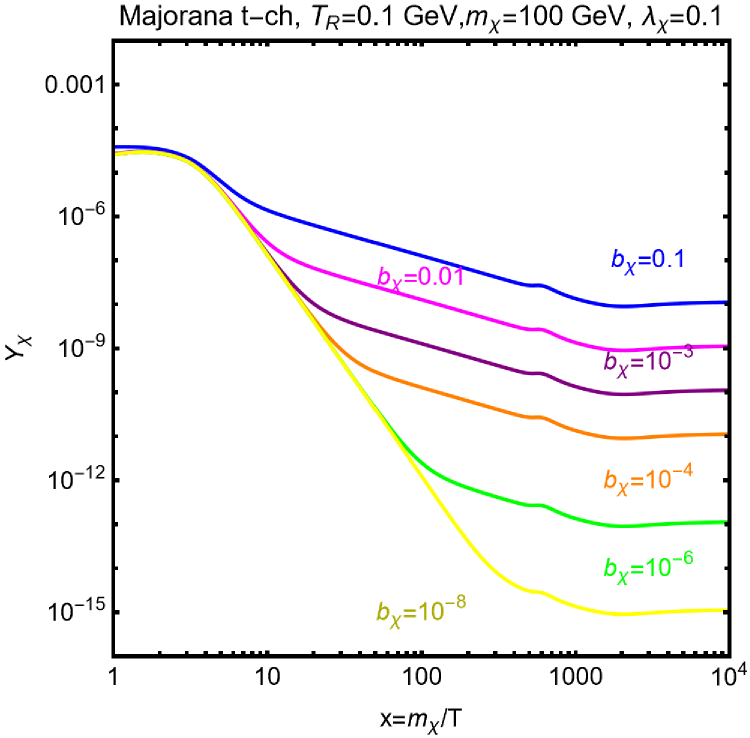}}
    \caption{\footnotesize{{\it Left Panel:} Non-thermal DM abundance for different values of $b_\chi$ and for $T_R=0.1\,\mbox{GeV}$ for a simplified model with DM belonging to a SU(2)-doublet and $m_\chi=100\,\mbox{GeV}$. {\it Right Panel}: The same but for a simplified model with $t$-channel mediator and Majorana DM.}}
    \label{fig:plotNTDMmath}
\end{figure}

To illustrate these distinct behaviors, we examine the time evolution of DM abundance for two specific model realizations, fixing the reheating temperature to $T_R = 0.1$ GeV. The results are presented in Fig.~\ref{fig:plotNTDMmath}.

For the SU(2)-doublet model with $m_\chi = 100$ GeV, the thermally averaged annihilation cross-section is approximately $10^{-24}\,\mbox{cm}^3\mbox{s}^{-1}$, which would yield a negligible relic density under conventional freeze-out. However, in the non-thermal production scenario, the correct relic density (corresponding to $Y_{\chi} \sim \mathcal{O}(10^{-11})$) is achieved in the reannihilation regime for $b_\chi \gtrsim 10^{-5}$.

The right panel of Fig.~\ref{fig:plotNTDMmath} examines the $t$-channel simplified model with Majorana DM, using parameters $\lambda_\chi=0.1$, $m_\chi=100\,\mbox{GeV}$, and $m_\Phi=3\,\mbox{TeV}$. Due to the $p$-wave dominated cross-section and additional suppression from the $m_\chi \ll m_\Phi$ mass hierarchy, condition \eqref{eq:reannihilation} is not satisfied. Consequently, the DM abundance far exceeds the experimentally favored value unless $b_\chi$ is particularly small, specifically requiring $b_\chi \in [10^{-6}, 10^{-5}]$.

\subsubsection{Direct Detection Prospects}

Having provided an insight about the solution of the Boltzmann's equation we can now investigate, in a more systematic way, the potential for direct detection (DD) facilities to probe the non-thermal DM production scenarios described above. We have hence repeated the parameter scans illustrated in the previous subsection but considering now a non zero $b_{\chi,\psi}$ varied in the $[10^{-10},1]$ range.


\begin{figure}
    \centering
    \subfloat{\includegraphics[width=0.5\linewidth]{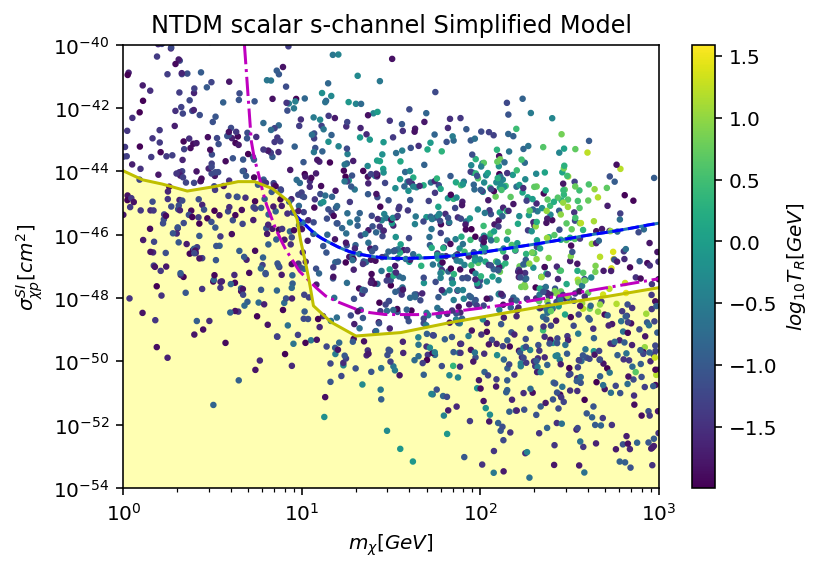}}
    \subfloat{\includegraphics[width=0.5\linewidth]{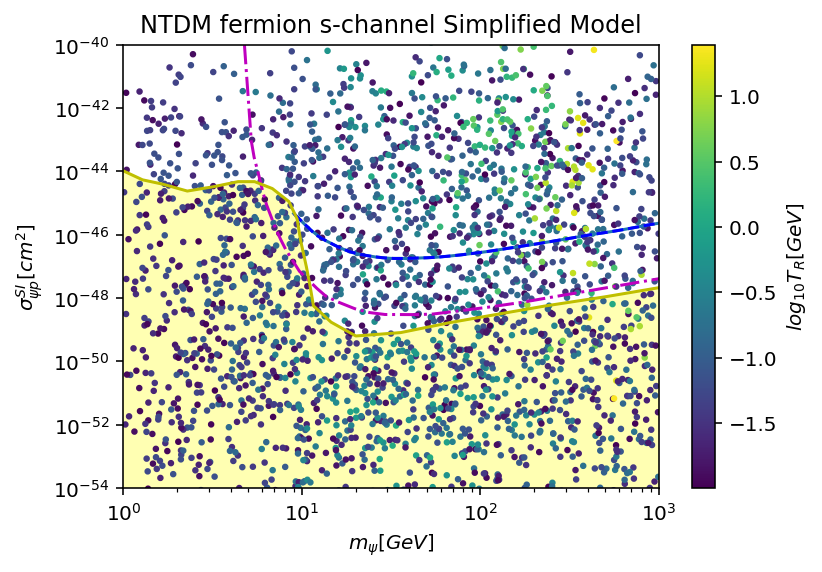}}\\
    \subfloat{\includegraphics[width=0.5\linewidth]{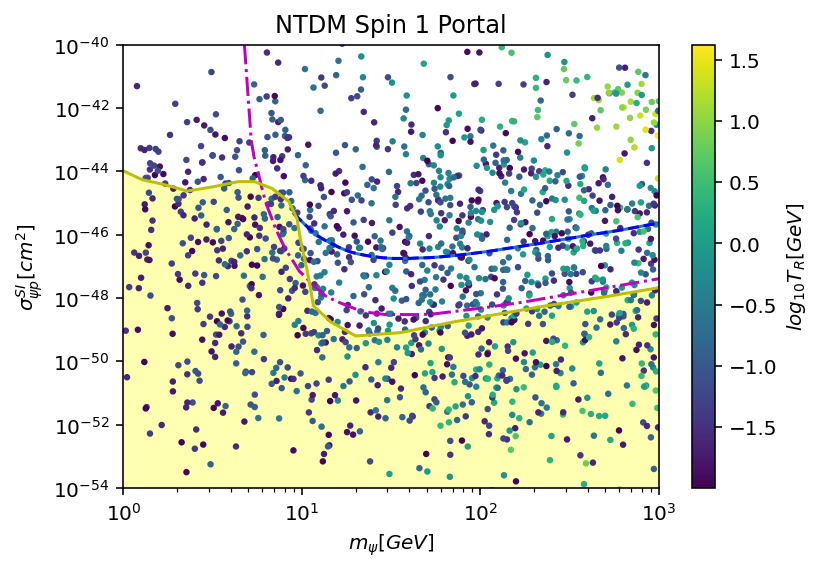}}
    \subfloat{\includegraphics[width=0.5\linewidth]{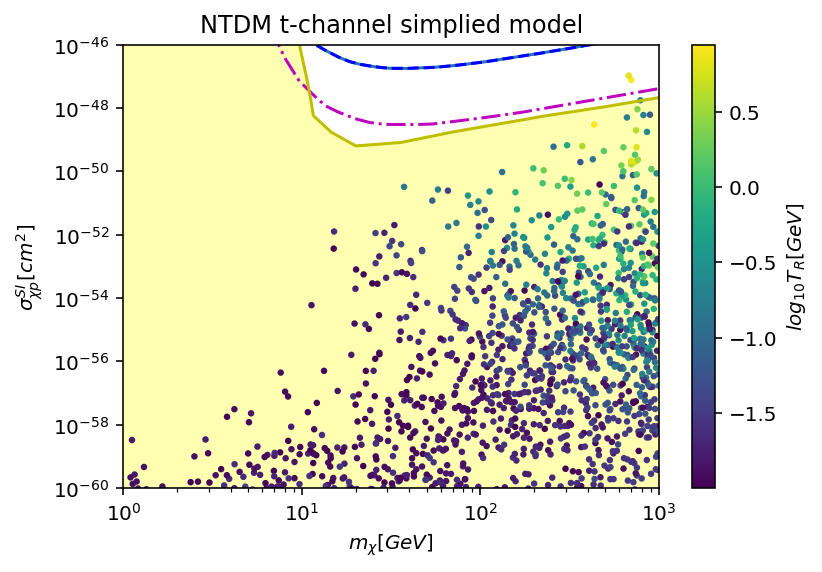}}
    \caption{\footnotesize{Parameter space viable for direct detection in non-thermal DM production scenarios. Results are shown for the simplified $s$-channel scalar mediator model (top left), $s$-channel fermion mediator model (top right), spin-1 portal model (bottom left), and $t$-channel mediator model (bottom right). Points are displayed in the $(m_\chi,\sigma_{\chi p}^{\rm SI})$ plane, with colors indicating the reheating temperature values.}}
    \label{fig:plotNTDMtt}
\end{figure}


Figure~\ref{fig:plotNTDMtt} displays the results in the $(m_\chi,\sigma_{\chi,\psi p}^{\rm SI})$ plane, with different colors corresponding to the reheating temperature values. As customary, the plots also include current exclusion limits from LZ, projected sensitivities of XLSD, and the neutrino floor region. As expected, accounting for non-thermal DM production significantly expands the parameter space accessible to direct detection experiments.

\begin{figure}
    \centering
    \subfloat{\includegraphics[width=0.5\linewidth]{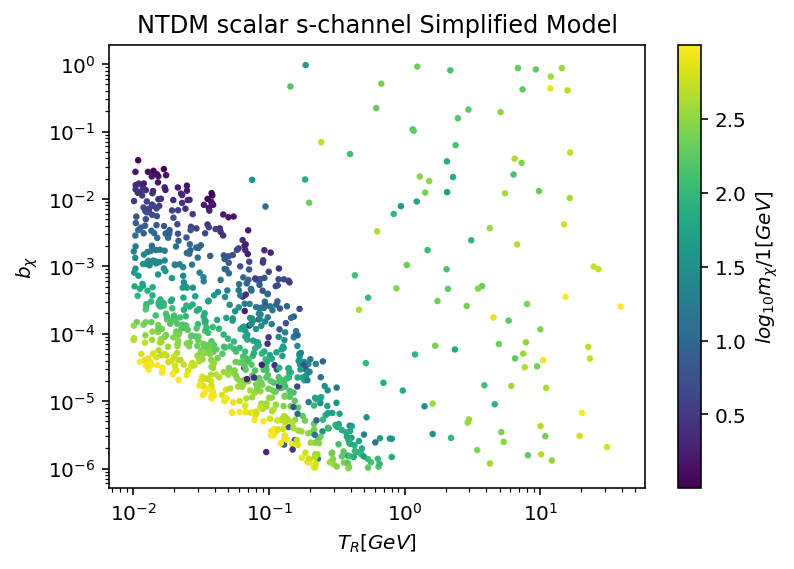}}
    \subfloat{\includegraphics[width=0.5\linewidth]{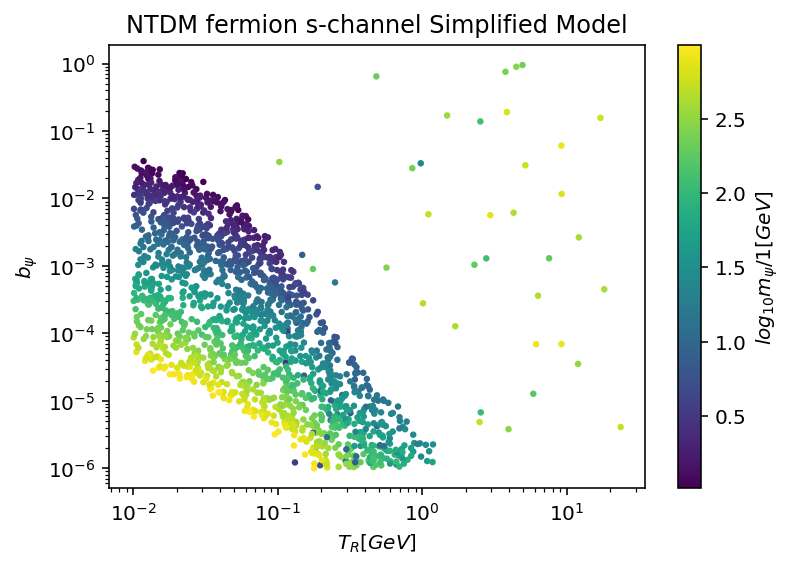}}\\
    \subfloat{\includegraphics[width=0.5\linewidth]{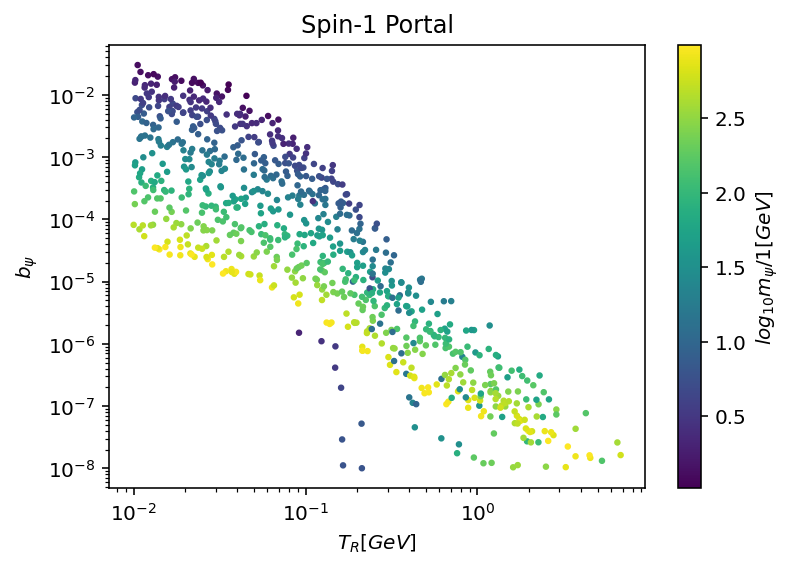}}\\
    \subfloat{\includegraphics[width=0.5\linewidth]{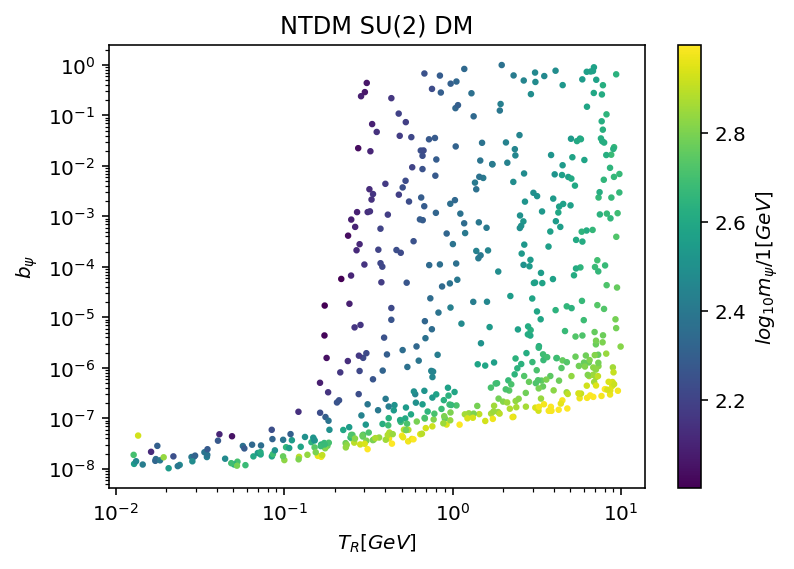}}
    \subfloat{\includegraphics[width=0.5\linewidth]{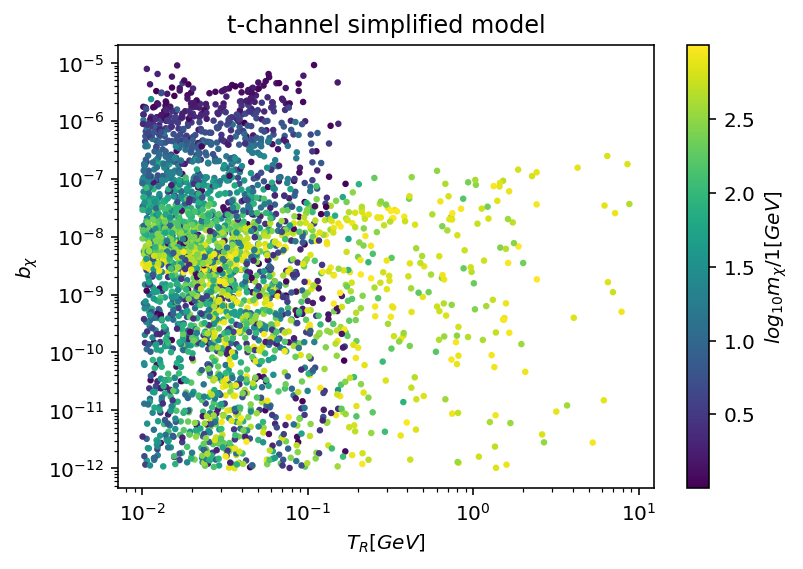}}
    \caption{\footnotesize{Model points for the simplified models satisfying both the correct DM relic density and current DD exclusion limits, assuming non-thermal production in an Early Matter dominated era. Points are displayed in the $(b_{\rm DM},T_R)$ plane, with colors indicating different DM mass values.}}
    \label{fig:pNTDMbTR}
\end{figure}

To better characterize the viable parameter space, Fig.~\ref{fig:pNTDMbTR} shows the model points that satisfy both the correct DM relic density and evade current LZ bounds, displayed in the $(b_{\rm DM},T_R)$ plane. The color coding tracks the DM mass values, revealing the complex interplay between the non-thermal production parameters and the resulting phenomenology. These results demonstrate that non-standard cosmological scenarios can substantially expand the discovery potential of direct detection experiments, motivating continued theoretical and experimental investigation of these alternative DM production mechanisms.


\subsection{Dark Matter with accelerated early expansion rates}



In this section, we explore the implications of modified cosmological histories on dark matter production, focusing on scenarios where the Universe undergoes accelerated expansion during the relevant epochs. We examine both the thermal freeze-out regime, where dark matter was initially in thermal equilibrium, and the freeze-in regime, where dark matter production occurs through feeble interactions. Our analysis demonstrates how non-standard cosmologies can reconcile theoretical predictions with observational constraints, particularly for models that face challenges in the standard cosmological framework.

Along this lines we hence consider the case in which the exotic component $\Phi$ can trigger an accelerated expansion of the Universe \cite{DEramo:2017gpl}:
\begin{equation}
    \rho_\Phi \propto a^{-(4+n)},\,\,\,\,n>0
\end{equation}
Among the various possibilities, quintessence \cite{Salati:2002md,Profumo:2003hq,Choi:1999xn} (corresponding to $n=2$, $\omega=-1$) is particularly popular.

Neglecting the case of non-thermal DM production, the non-standard cosmological history affects the freeze-out of the DM, setting it at an earlier stage with respect to standard cosmology. The two temperatures are related by the following approximate equation:

\begin{equation}
    x_{s.f.o.}e^{-x_{s.f.o.}}=x_{f.o.}e^{-x_{f.o}}{\left(\frac{x_{f.o.}}{x_r}\right)}^{n/2}
\end{equation}
where $x_{s.f.o}$ corresponds to the freeze-out in a standard (i.e., radiation-dominated) cosmological history while $x_r=m_\chi/T_r$. The temperature $T_r$ represents a reference temperature marking the end of the non-standard cosmological history. It is defined as the temperature at which the energy densities of radiation and $\Phi$ coincide, so that for $T< T_r$ the Universe returns to being radiation-dominated:
\begin{equation}
    H=\sqrt{\frac{\rho(T)}{3 M_{\rm Pl}^2}}\,\,\,\,\,\,\rho(T)=\rho_R (T)\left[1+\frac{g_{*}(T_r)}{g_{*}(T)}{\left(\frac{g_{*,s}(T)}{g_{*,s}(T_r)}\right)}^{(4+n)/3}{\left(\frac{T}{T_r}\right)}^n\right]
\end{equation}

Semi-analytical solutions of the Boltzmann equations can be obtained in the case $n\geq 2$. Considering an $s$-wave dominated annihilation cross-section $\langle \sigma v \rangle=\sigma(s=4 m_\chi^2)=a$, the DM abundance can be approximately written as:
\begin{align}
    & Y_\chi (x)\simeq \frac{x_r}{m_\chi M_{pl} a}{\left[\frac{2}{x_{f.o.}}+\log \left(\frac{x}{x_{f.o.}}\right)\right]}^{-1},\,\,\,\,n=2\nonumber\\
    & Y_{\chi}(x)\simeq \frac{x_r^{n/2}}{2m_\chi M_{pl}a}{\left[x_{f.o}^{n/2-2}+\frac{x^{n/2-1}}{n-2}\right]}^{-1},\,\,\,\,n>2
\end{align}

\begin{figure}
    \centering
    \includegraphics[width=0.6\linewidth]{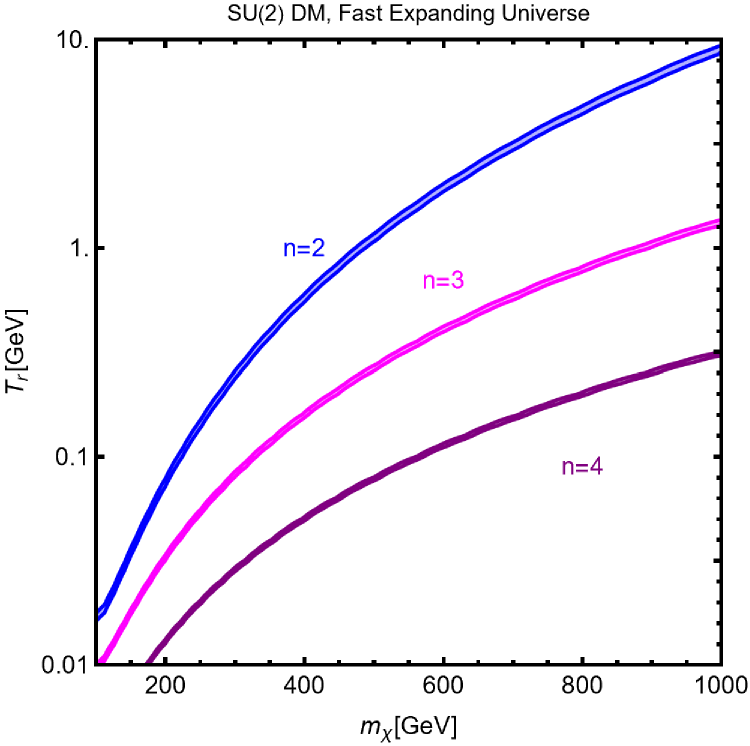}
    \caption{\footnotesize{Parameter space for SU(2)-doublet dark matter in non-standard cosmological scenarios. The plot shows the viable region in the $(m_\chi, T_r)$ plane where the correct relic density can be achieved through thermal freeze-out in a fast-expanding Universe.}}
    \label{fig:prelentless1}
\end{figure}

One of the key aspects of thermal freeze-out during accelerated expansion of the Universe is that the DM relic density is enhanced with respect to standard cosmology. Higher annihilation cross-sections are needed to match the experimentally favored value. This setup offers a further solution to accommodate the correct relic density for SU(2)-multiplet DM with masses below the TeV scale. We have therefore determined the viable parameter space for the simplified model with SU(2)-doublet DM and shown in Fig.~(\ref{fig:prelentless1}).

\begin{figure}
    \centering    
    \subfloat{\includegraphics[width=0.4\linewidth]{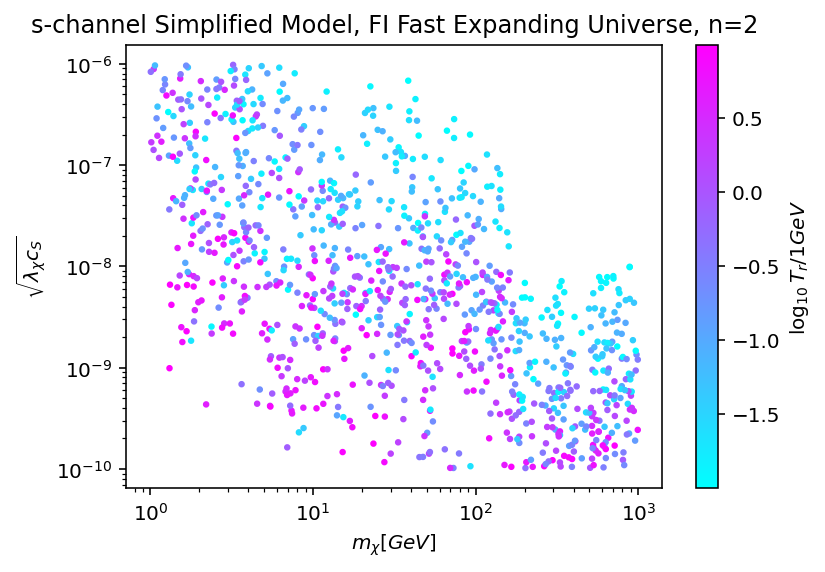}}
    \subfloat{\includegraphics[width=0.4\linewidth]{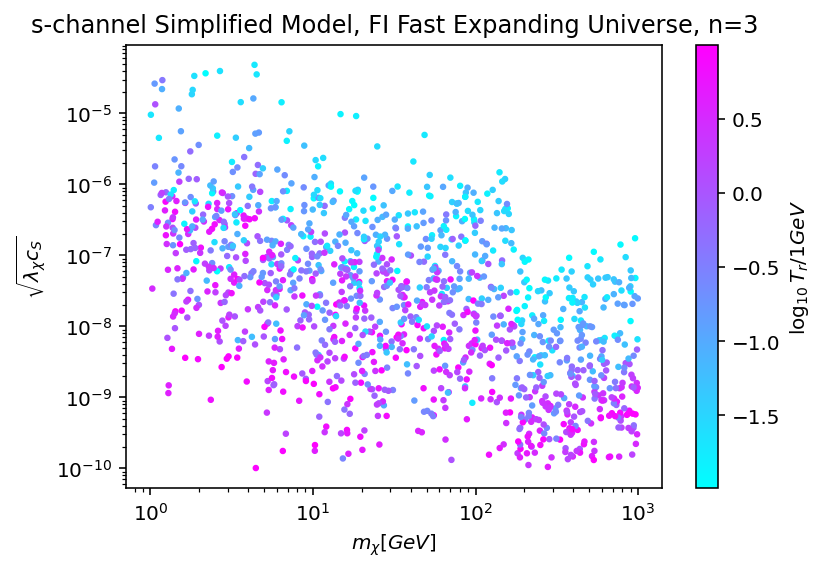}}
    \subfloat{\includegraphics[width=0.4\linewidth]{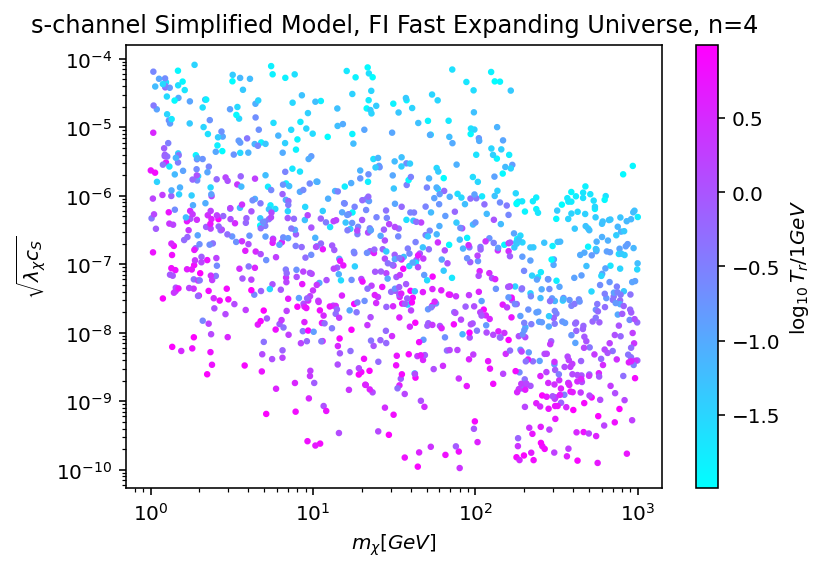}}\\
    \subfloat{\includegraphics[width=0.4\linewidth]{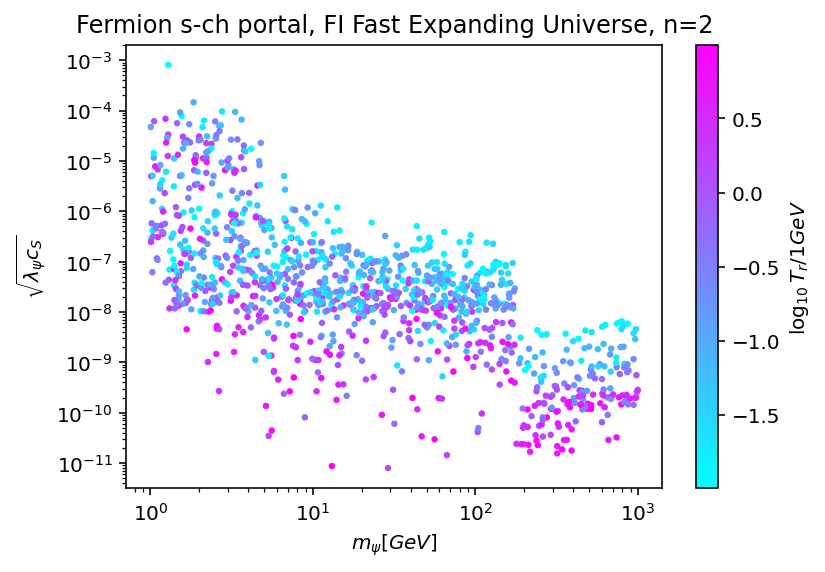}}
    \subfloat{\includegraphics[width=0.4\linewidth]{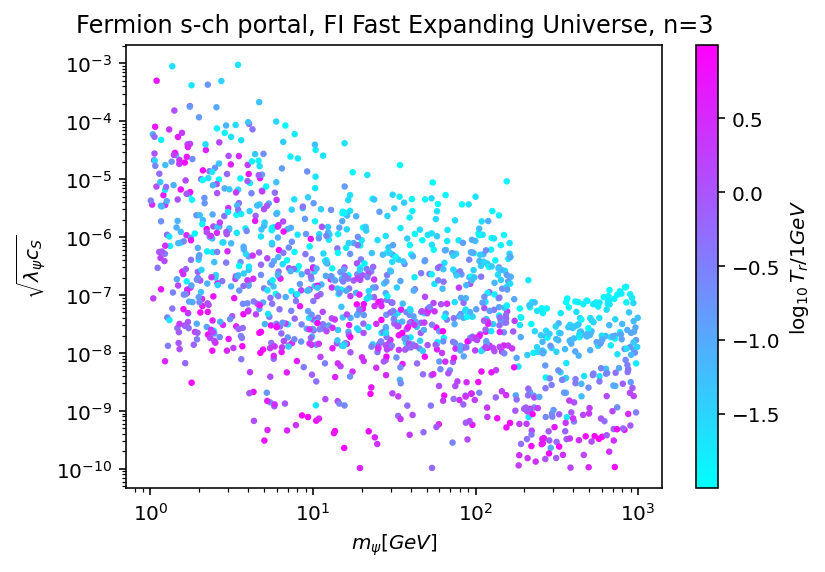}}
    \subfloat{\includegraphics[width=0.4\linewidth]{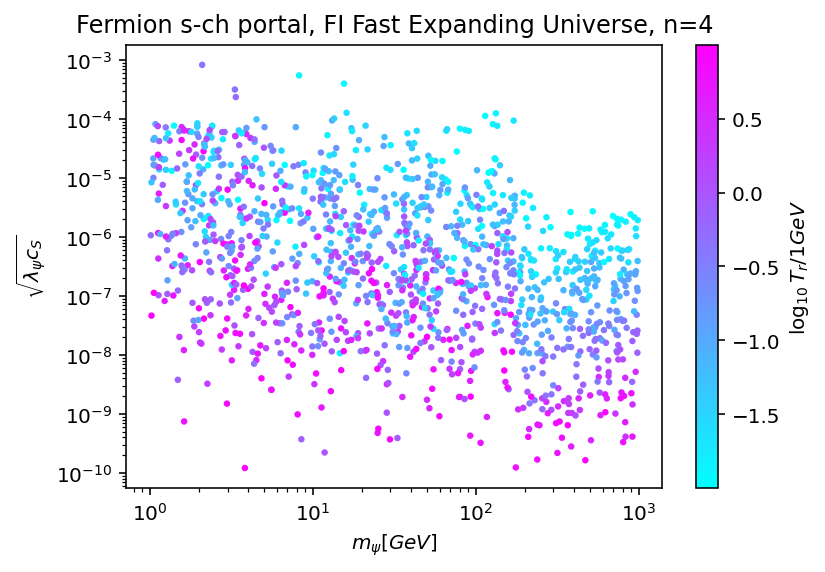}}\\
    \subfloat{\includegraphics[width=0.4\linewidth]{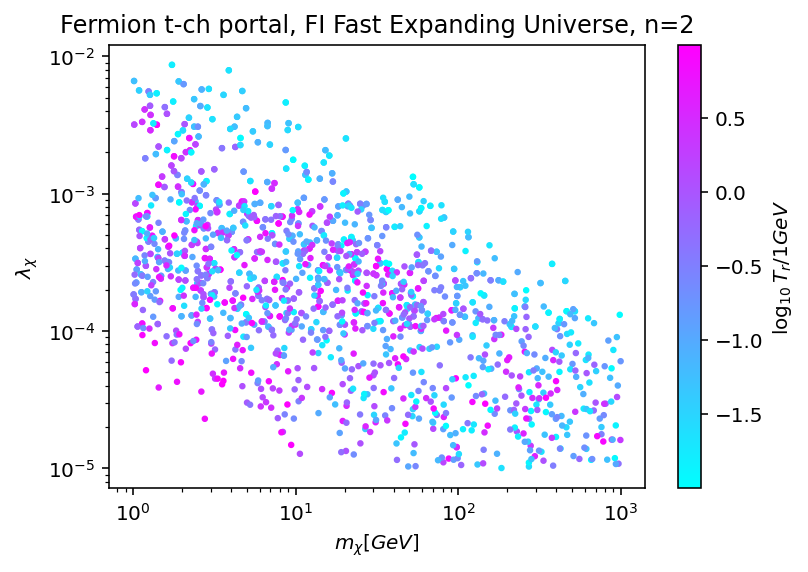}}
    \subfloat{\includegraphics[width=0.4\linewidth]{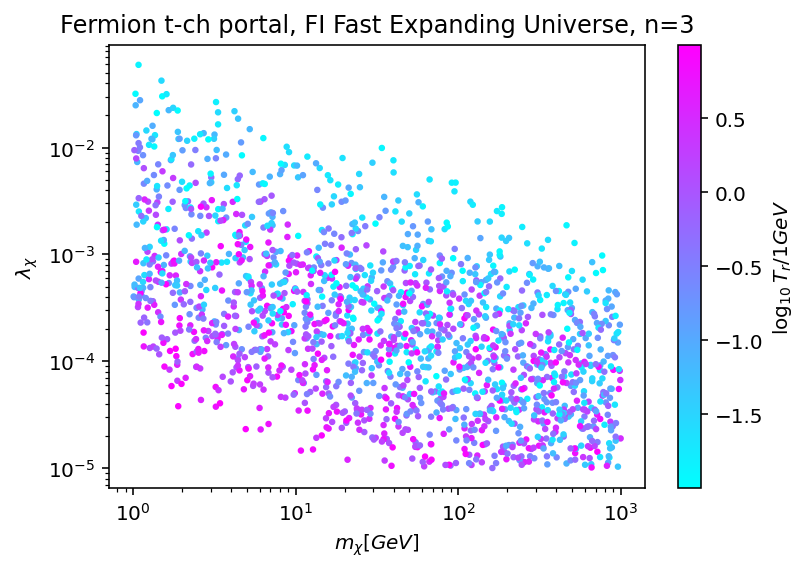}}
    \subfloat{\includegraphics[width=0.4\linewidth]{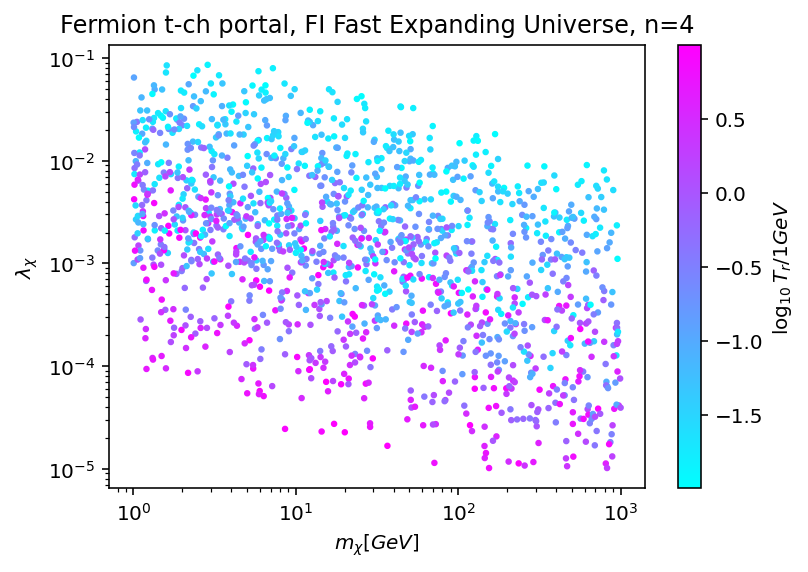}}
    \caption{\footnotesize{Relic density of feebly interacting DM in the fast expanding regime. The case of study is represented by the $s$-channel simplified model with scalar DM (top panels), fermionic DM (middle panels) and $t$-channel simplified models (bottom panels). Each panel shows the outcome of a parameter scan (see main text for details). The three different panels correspond, from left to right, to the cases $n=2,3,4$. The panels show the model points with the correct DM relic density in the $(m_\chi,\sqrt{\lambda_\chi^S c_S}/\sqrt{g_\psi c_S}/\lambda_\chi)$ bidimensional plane with a color pattern tracking $T_r$.}}
    \label{fig:plot_relentless}
\end{figure}

The figure shows, in the $(m_\chi,T_r)$ parameter space, the isocontours corresponding to the correct relic density for $n=2$ (blue curve), $n=3$ (magenta curve) and $n=4$ (purple curve).

The case of a fast-expanding Universe is also of particular interest for feebly interacting particles. As evidenced in \cite{DEramo:2017ecx}, a general analytical solution cannot be achieved as different processes can contribute to the generation of the relic density. A general feature is nevertheless that DM production is suppressed with respect to a radiation-dominated Universe; consequently, comparatively higher couplings are needed. This feature is particularly interesting as it would potentially put a freeze-in scenario within the reach of next-generation detectors.
We provide a (semi)analytical insight by considering, as case of study the $s$-channel simplified model. A similar reasoning can be applied to the other portal models. The freeze-in regime can be achieved for this model by considering the regime $m_\chi \ll m_S$ and very small values of the $\lambda_\chi^S,c_S$ parameters. In this setup one can straightforwardly adapt the expression provided in \cite{DEramo:2017ecx} for the case of pair production of DM from annihilations of SM states by taking:
\begin{equation}
    \left \vert M_{SM SM \rightarrow \chi \chi}\right \vert^2=(\lambda_\chi^S)^2 c_S^2 \frac{m_f^2}{v_h^2}\frac{m_\chi^4}{m_S^4}
\end{equation}
so that one obtains:
\begin{equation}
    Y_\chi(x)\simeq \frac{135\sqrt{10}}{256\pi^8}\frac{(\lambda_\chi^S)^2 c_S^2}{g_{*}^{3/2}}\frac{m_\chi^3 M_{Pl}}{m_S^4}\frac{m_f^2}{v_h^2}x_r^{-n/2}\int_0^x dx^{'}K_1(x^{'})x^{'\,2+n/2}
\end{equation}

Fig.~(\ref{fig:plot_relentless}) shows the outcome of a parameter scan of the simplified $s$-channel (both scalar and fermionic DM) and $t$-channel models over the following ranges:
\begin{align}
    & m_{\chi}\in \left[1,1000\right]\,\mbox{GeV},\,\,\,\,m_S\in [1,10]\,\mbox{TeV},\,\,\,\sqrt{c_S \lambda_\chi^S}\in \left[10^{-10},10^{-3}\right]\nonumber\\
    & m_{\psi}\in \left[1,1000\right]\,\mbox{GeV},\,\,\,\,m_S\in [1,10]\,\mbox{TeV},\,\,\,\sqrt{c_S g_\psi}\in \left[10^{-10},10^{-3}\right]\nonumber\\
    & m_{\chi}\in \left[1,1000\right]\,\mbox{GeV},\,\,\,\,m_\Phi\in [1,10]\,\mbox{TeV},\,\,\,\lambda_\chi\in \left[10^{-5},10^{-1}\right]\nonumber\\
\end{align}
In each case the temperature associated to the end of the non-standard cosmological phase has been varied as $T_r \in \left[10\,\mbox{MeV},10\,\mbox{GeV}\right]$ with, as customary, the implicit requirement $T_r < T_{s.f.o.}$.

The figure shows the model points featuring the correct DM relic density in the $(m_\chi,\sqrt{\lambda_\chi^S c_S},\sqrt{g_\psi c_S},\lambda_\chi)$ parameter space, following a color pattern determined by the value of $T_r$. The results demonstrate that non-standard cosmological histories can significantly modify the viable parameter space for dark matter models, opening new regions that would be excluded in standard cosmology. This is particularly relevant for models with weak couplings, where the enhanced expansion rate during the non-standard epoch can compensate for the reduced interaction strength, allowing these models to achieve the correct relic density while remaining compatible with experimental constraints.

The implications of these results extend beyond the specific models considered here. Non-standard cosmological histories provide a general framework for reconciling dark matter theory with observation, offering new pathways to the observed relic density that can evade current experimental bounds. This is particularly important for models that face the ``WIMP miracle problem'', where the cross-sections required for thermal freeze-out are in tension with direct detection limits. By modifying the cosmological context, we can maintain the elegance of thermal production while avoiding the most stringent experimental constraints.

\section{Conclusion}

The quest to understand the nature of dark matter has reached a critical juncture as direct detection experiments approach the fundamental neutrino floor limit. Our comprehensive analysis of theoretical benchmarks for ultimate WIMP detectors reveals both the challenges and opportunities that lie ahead in this final phase of the traditional dark matter search paradigm.

\subsection{Theoretical Landscape at the Neutrino Floor}

Our systematic examination of simplified dark matter models demonstrates that the parameter space accessible to next-generation experiments remains remarkably rich, even in the face of the irreducible neutrino background. The interplay between thermal freeze-out and freeze-in production mechanisms opens multiple avenues for discovery, with each pathway offering distinct phenomenological signatures that can guide experimental strategies. Particularly compelling are scenarios involving momentum-dependent interactions and blind-spot configurations, which naturally suppress direct detection cross-sections while maintaining cosmologically viable relic densities.

The exploration of electroweakly interacting multiplets and $t$-channel mediator models reveals that loop-induced effects often dominate the detection prospects for theoretically well-motivated scenarios~\cite{Hisano:2011cs,Arcadi:2024abc}. This finding underscores the critical importance of higher-order calculations in establishing reliable theoretical predictions for ultimate sensitivity experiments. Our analysis shows that many models previously considered beyond experimental reach may become accessible through careful consideration of radiative corrections and multi-loop contributions.

For the $s$-channel scalar portal model, the relationship between relic density and direct detection sensitivity is governed by the thermally averaged annihilation cross-section:
\begin{equation}
\langle\sigma v\rangle(\chi\chi \to \bar{f}f) \approx \sum_f \frac{3n_f^c}{16\pi} \frac{(\lambda_\chi^S)^2 c_S^2 m_f^2}{v_h^2} \frac{m_\chi^2}{(4m_\chi^2 - m_S^2)^2} \left[1 - \frac{4m_f^2}{m_\chi^2}\right]\theta(m_\chi - m_f)
\end{equation}
which translates to a predicted spin-independent scattering cross-section of:
\begin{equation}
\sigma_{\chi p}^{\text{SI}} \approx 1.1 \times 10^{-42}\text{ cm}^2 \left(\frac{100\text{ GeV}}{m_\chi}\right)^4 \left(\frac{0.12}{\Omega_\chi h^2}\right)^{-1}
\end{equation}

\subsection{Non-Standard Cosmological Scenarios}

One of the most significant insights from our work concerns the dramatic impact of non-standard cosmological histories on the relationship between relic abundance and direct detection prospects. Early matter domination scenarios and fast-expanding Universe models can fundamentally alter the traditional freeze-out dynamics, enabling viable dark matter candidates with properties that differ substantially from conventional expectations~\cite{Arias:2019abc,DEramoFernandez:2017abc,SilvaMalpartida:2024abc}.

In the early matter domination regime, the dark matter yield satisfies the modified Boltzmann equation system with the critical condition for reannihilation:
\begin{equation}
\langle\sigma v\rangle > 1.96 \times 10^{-29}\text{ cm}^3\text{s}^{-1} \, b_{\text{DM}}^{-1} \left(\frac{1\text{ GeV}}{T_R}\right)^{4/3}
\end{equation}
where \(T_R\) is the reheating temperature and \(b_{\text{DM}}\) parameterizes the branching ratio for dark matter production from the exotic component decay.

These mechanisms provide natural explanations for suppressed direct detection signals while maintaining the correct cosmological abundance, effectively extending the viable parameter space well into regions that will be probed by ultimate sensitivity experiments. The freeze-in paradigm emerges as particularly promising in this context, as it naturally accommodates the ultra-weak couplings characteristic of models approaching the neutrino floor.

For freeze-in production in fast-expanding cosmologies with equation of state parameter \(\omega = -1 - n/3\), the dark matter yield is approximately:
\begin{equation}
Y_\chi(x) \simeq \frac{135\sqrt{10}}{256\pi^8} \frac{(\lambda_\chi^S)^2 c_S^2}{g_*^{3/2}} \frac{m_\chi^3 M_{\text{Pl}}}{m_S^4} \frac{m_f^2}{v_h^2} x_r^{-n/2} \int_0^x dx' K_1(x') x'^{2+n/2}
\end{equation}
where \(x_r = m_\chi/T_r\) marks the transition temperature.

\subsection{Experimental Strategies and Complementarity}

The approach to the neutrino floor necessitates a fundamental shift in experimental strategy, moving beyond simple sensitivity improvements toward sophisticated background discrimination techniques~\cite{Billard:2013cfa,OHare:2020wah}. Our benchmarks highlight the critical importance of multi-target approaches, directional sensitivity, and statistical separation methods in breaking the degeneracy between neutrino-induced and WIMP-induced nuclear recoils. The complementarity between different experimental approaches—direct detection, indirect detection, and collider searches—becomes increasingly crucial as individual techniques approach their fundamental limits.

Current experiments such as LUX-ZEPLIN have achieved sensitivities approaching \(10^{-47}\text{ cm}^2\) for spin-independent WIMP-nucleon scattering~\cite{Akerib:2022ort}, while next-generation detectors like DARWIN/XLZD promise to extend this reach to \(\sim 10^{-48}\text{ cm}^2\). The neutrino floor represents a fundamental barrier where coherent elastic neutrino-nucleus scattering creates an irreducible background that mimics WIMP signals.

Emerging technologies, including quantum sensing approaches and gravitational wave detectors, offer unprecedented opportunities to probe previously inaccessible parameter regions. These novel detection strategies can potentially circumvent the traditional neutrino floor limitations by targeting different interaction mechanisms or exploiting ultra-high precision measurements that were previously impossible.

\subsection{Beyond the WIMP Paradigm}

While our analysis focuses primarily on WIMP-like candidates, the broader implications extend to the entire landscape of dark matter theories. The recognition that traditional thermal freeze-out models may be fundamentally limited by neutrino backgrounds has catalyzed renewed interest in alternative production mechanisms and non-minimal dark sectors~\cite{Cosme:2023bkx,Hambye:2018dpi}. Following an analogous approach as for example \cite{Arcadi:2024abc,Arcadi:2025sxc}, we will consider DM masses in the $1\,\mbox{GeV}-1\,\mbox{TeV}$ range.

For electroweakly interacting dark matter multiplets, the Lagrangian takes the form:
\begin{equation}
\mathcal{L} = \frac{g_2}{4}[n^2 -(2Y + 1)^2]\chi\gamma^\mu\psi^-W^+_\mu + \frac{g_2}{4}[n^2 -(2Y -1)^2]\chi\gamma^\mu\psi^+W^-_\mu -i\frac{g_2Y}{\cos\theta_W}\chi\gamma^\mu\eta Z_\mu
\end{equation}
where \(n\) represents the SU(2) multiplet dimension and \(Y\) the hypercharge assignment~\cite{Cirelli:2005uq,Cirelli:2014nga}.

Multi-messenger approaches, combining information from gravitational waves, gamma-ray observations, and neutrino detections, promise to open entirely new windows into the dark sector. The development of comprehensive theoretical frameworks enables systematic exploration of models with similar phenomenological predictions, which will prove essential for interpreting potential signals in the complex landscape near the neutrino floor.

\subsection{Future Prospects and Roadmap}

Looking toward the next decade, several key developments will shape the future of dark matter searches at the neutrino floor. The deployment of next-generation multi-ton detectors will provide unprecedented sensitivity to spin-independent WIMP-nucleon interactions, potentially reaching cross-sections as low as \(10^{-48}\text{ cm}^2\). Simultaneously, advances in quantum sensing technology and novel detection concepts will enable exploration of previously inaccessible mass ranges, from sub-GeV to ultra-light axion-like particles.

The relic density constraint in the standard freeze-out scenario is given by:
\begin{equation}
\Omega_\chi h^2 \simeq 1.07 \times 10^9\text{ GeV}^{-1} \frac{x_f}{g_*^{1/2}} \frac{1}{M_{\text{Pl}}(a + 3b/x_f)}
\end{equation}
where \(x_f = m_\chi/T_f\) is the freeze-out parameter, \(g_*\) the effective number of relativistic degrees of freedom, and \(a\), \(b\) parameterize the $s$-wave and $p$-wave contributions to the velocity expansion.

The theoretical framework developed in this work provides essential guidance for optimizing these experimental efforts. Our benchmarks identify the most promising parameter regions for discovery, highlight the importance of complementary detection strategies, and establish the foundation for interpreting results as experiments push into uncharted territory. The systematic exploration of non-standard cosmological scenarios and alternative production mechanisms ensures that the theoretical landscape remains rich with possibilities, even as traditional WIMP searches approach their fundamental limits.

\subsection{Final Remarks}

The ultimate success of dark matter detection efforts will likely require a coordinated approach combining multiple experimental strategies, sophisticated theoretical modeling, and innovative technologies that push beyond current paradigms. Our analysis demonstrates that the neutrino floor, rather than representing an insurmountable barrier, marks the beginning of a new era in dark matter searches—one characterized by unprecedented precision, novel detection concepts, and a deeper understanding of the rich theoretical landscape that connects particle physics, cosmology, and fundamental interactions.

The WIMP miracle relation \(\langle\sigma v\rangle \sim 3 \times 10^{-26}\text{ cm}^3/\text{s}\) continues to provide theoretical motivation, even as experimental constraints push models toward increasingly complex scenarios involving loop-induced interactions, non-thermal production mechanisms, and exotic cosmological histories~\cite{Kahlhoefer:2015vua,Arcadi:2017jse}.

The journey toward understanding the nature of dark matter continues, with the neutrino floor serving not as a destination, but as a waypoint toward even more ambitious goals. The theoretical benchmarks established in this work provide a roadmap for this continuing quest, ensuring that the next generation of experiments will be optimally positioned to make the discoveries that will finally illuminate one of the universe's deepest mysteries. As we approach the limits of traditional detection methods, the interplay between theoretical innovation and experimental ingenuity will prove crucial in maintaining momentum toward the ultimate goal of dark matter discovery and characterization.


\section*{Acknowledgments}

G.A. acknowledges support from the DAAD German Academic Exchange Service. G.A. thanks the MPIK for the warm hospitality during part of the completion of this work. This work is partly supported by the U.S.\ Department of Energy grant number de-sc0010107 (SP).

\appendix

\bibliographystyle{utphys}
\bibliography{biblio}
\end{document}